\documentclass[%
 reprint,
 amsmath,amssymb,
 aps,
prc
floatfix,
]{revtex4-2}

\usepackage{graphicx}
\usepackage{dcolumn}
\usepackage{bm}

\begin{document}

\preprint{APS/123-QED}

\title{Low-lying level structures in $^{162}$Lu}

\author{N Susshma, Deepa Seetharaman, K Vijay Sai}
\author{R Gowrishankar}%
 \email{rgowrishankar@sssihl.edu.in}
\affiliation{%
 Department of Physics, Sri Sathya Sai Institute of Higher Learning, Prasanthi Nilayam-515134, India
}%

\date{\today}

\begin{abstract}
\textbf{Background:} The low-lying level structure of deformed odd-odd nuclei away from the region of stability have been sparsely characterized due to low cross-section yields in reactions, short half-lives and lack of dedicated studies. One such interesting case is the odd-odd rare earth light-mass isotope $^{162}$Lu, where except for the ground state (gs) J$^{\pi}$=1$^-$\{$\pi$1/2$^+$[411$\downarrow$]$\otimes$$\nu$3/2$^-$[521$\uparrow$]\}, assignments of other low-lying levels are unknown or ambiguous. The nucleus is reported to have two low-lying isomeric states with  ambiguous assignments of spin-parity, level energy, orbital configuration without specific information on the individual $\beta$-decay branchings. 

\textbf{Purpose:} This work uses the well-tested empirical Two Quasiparticle Rotor Model (TQRM) to deduce the low-energy spectrum of $^{162}$Lu and thereby characterize the isomeric states. Following the characterization, similar trend of isomers in neighboring odd-odd  $^{164,166,168}$Lu isotopes are analyzed using the empirical model.  

\textbf{Methods:} The empirical three-step TQRM calculations estimate the spin-parities, orbital configurations, and energies of the physically admissible two quasiparticle (2qp) states in the odd-odd system. Experimental energies of the single particle (1qp) proton and neutron states close to the Fermi surface in the mass region of interest are taken as inputs. The spin-parities of the low-lying 2qp bandheads are estimated by coupling the 1qp orbitals and the energies are calculated using the empirical formula. Beta decay selection rules are employed for mapping the favored beta branching of the gs and isomers to select daughter levels.  

\textbf{Results:} The 1.5 min isomer in $^{162}$Lu is confirmed to have J$^\pi$= 4$^-$\{$\pi$5/2$^+$[402$\uparrow$] $\otimes$  $\nu$3/2$^-$[521$\uparrow$]\}, with estimated level energy  E$_x$ $\thickapprox$ 62 keV. The 1.9 min isomer is characterized for the first time with J$^\pi$= 6$^+$\{$\pi$9/2$^-$[514$\uparrow$] $\otimes$ $\nu$3/2$^-$[521$\uparrow$]\} and energy E$_x$ $\thickapprox$ 157 keV. Based on these assignments, individual $\beta$-decay branches of ground state and the two isomers to the daughter levels in $^{162}$Yb are proposed. The study is extended to neighboring isotopes $^{164,166,168}$Lu. A short-lived isomer, tentatively proposed in a previous study of $^{164}$Lu, is identified as J$^\pi$= 4$^-$\{$\pi$5/2$^+$[402$\uparrow$] $\otimes$ $\nu$3/2$^-$[521$\uparrow$]\} with energy E$_x$ $\thickapprox$ 52 keV. The energy of the J$^\pi$=3$^+$ isomer in $^{168}$Lu, currently listed with large uncertainty, is deduced as E$_x$ $\thickapprox$ 100 keV significantly improving its value.  

\textbf{Conclusions:} The low-lying levels of $^{162}$Lu were constructed using the empirical TQRM calculations. The spin-parities, orbital configurations and level energies of the two low-lying isomers have been estimated and their beta decay pathways are proposed for the first time. Isomers in the neighboring odd-odd isotopes have also been studied. These proposed assignments can be used as location guides for future experimental studies.

\end{abstract}

\maketitle


\section{\label{sec1}Introduction}

The odd-odd light-mass Lu isotope $^{162}$Lu was first experimentally produced and reported along with its isotopic neighbors $^{164,166}$Lu by Neiman and Ward \cite{NW}. The $\beta$-decay studies of $^{162}$Lu, that were subsequently carried out in various works, are the source of its existing low-lying level structure \cite{ENSDF}. 
The adopted scheme has been derived from two important decay studies, namely, Schrewe \textit{et al.} \cite{Schrewe} and Hild \textit{et al.} \cite{Hild162}. Both studies carried out decay spectroscopy of $^{162}$Hf with the objective of comparing the percentage of $\alpha$- and $\beta$-decay in neutron-deficient Hf isotopes. Through detailed study of $\gamma$-$\gamma$ coincidences, Schrewe \textit{et al.} \cite{Schrewe}, placed four energy levels, including the ground state (gs) based on three intense $\gamma$-transitions observed following the $\beta$-decay of $^{162}$Hf. The decay spectroscopy by Hild \textit{et al.} \cite{Hild162} confirmed these reported $\gamma$-rays adding eight new transitions in coincidence, (possibly) belonging to $^{162}$Lu and proposed the low-energy spectrum of $^{162}$Lu up to E$_x$ $\thickapprox$ 600 keV. However, this adopted level scheme defines only the energies of the levels based on the observed $\gamma$ transitions with no spin-parity (J$^\pi$) or orbital configuration assignments. There has been no dedicated study otherwise to unambiguously characterize the low-lying level structure of $^{162}$Lu. In contrast, its high-spin rotational  structures, including triaxially super deformed bands have been extensively analyzed and reported by multiple high-energy reaction studies \cite{Cardona97,Gupta97,Bringel03}. The first  experimental observation of wobbling bands was reported in $^{163}$Lu \cite{163Lu} after which it was reported in neighboring $^{161,162,164,165,167}$Lu \cite{161Lu,Bringel03,164LuTMT,164Lutriax,165Lu,167Lu,Lu_triax} isotopes through experimental and theoretical studies. These nuclei are hence regions of exotic shapes and structure deformations. Characterizing their low-lying level structures would add to the comprehensive understanding of these known realms of exotic properties.  

The spin-parity of the $^{162}$Lu gs was determined to be J$^\pi$=1$^-$ with orbital configuration  $\pi$1/2[411$\downarrow$] $\otimes$$\nu$3/2[521$\uparrow$] through collinear laser spectroscopy \cite{Georg98}. The orbital configuration was assigned by comparing the calculated and experimental magnetic moments. The earlier decay studies by Burman \textit{et al.} \cite{Burman} and Alkhazov \textit{et al.} \cite{Alkhazov} determined the half-life of the gs as t$_{1/2}$$\thickapprox$1.4 min. The current ENSDF \cite{ENSDF} lists two short-lived isomeric states, one with a tentative  J$^\pi$=(4$^-$) and half-life t$_{1/2}$= 1.5 min and another with undetermined spin-parity and t$_{1/2}$=1.9 min \cite{ENSDF}. These isomeric states were reported in a lone decay study by Behrens \cite{Behrens} and have not been subjected to further characterization. The gs and two isomers are expected to undergo independent  $\beta$-decays ($\beta^+$+$\epsilon$) to the levels of the daughter nucleus $^{162}$Yb \cite{Alkhazov,Burman,Schrewe,Hild162}. 
The combined decays of the gs and the two isomers to the levels of $^{162}$Yb are described in the recently updated Nuclear Data Sheets (NDS) \cite{NDS}, while the individual $\beta$-branches  remain undetermined. 

\begin{figure*}[!t]
\centering
\includegraphics[width=18.0cm]{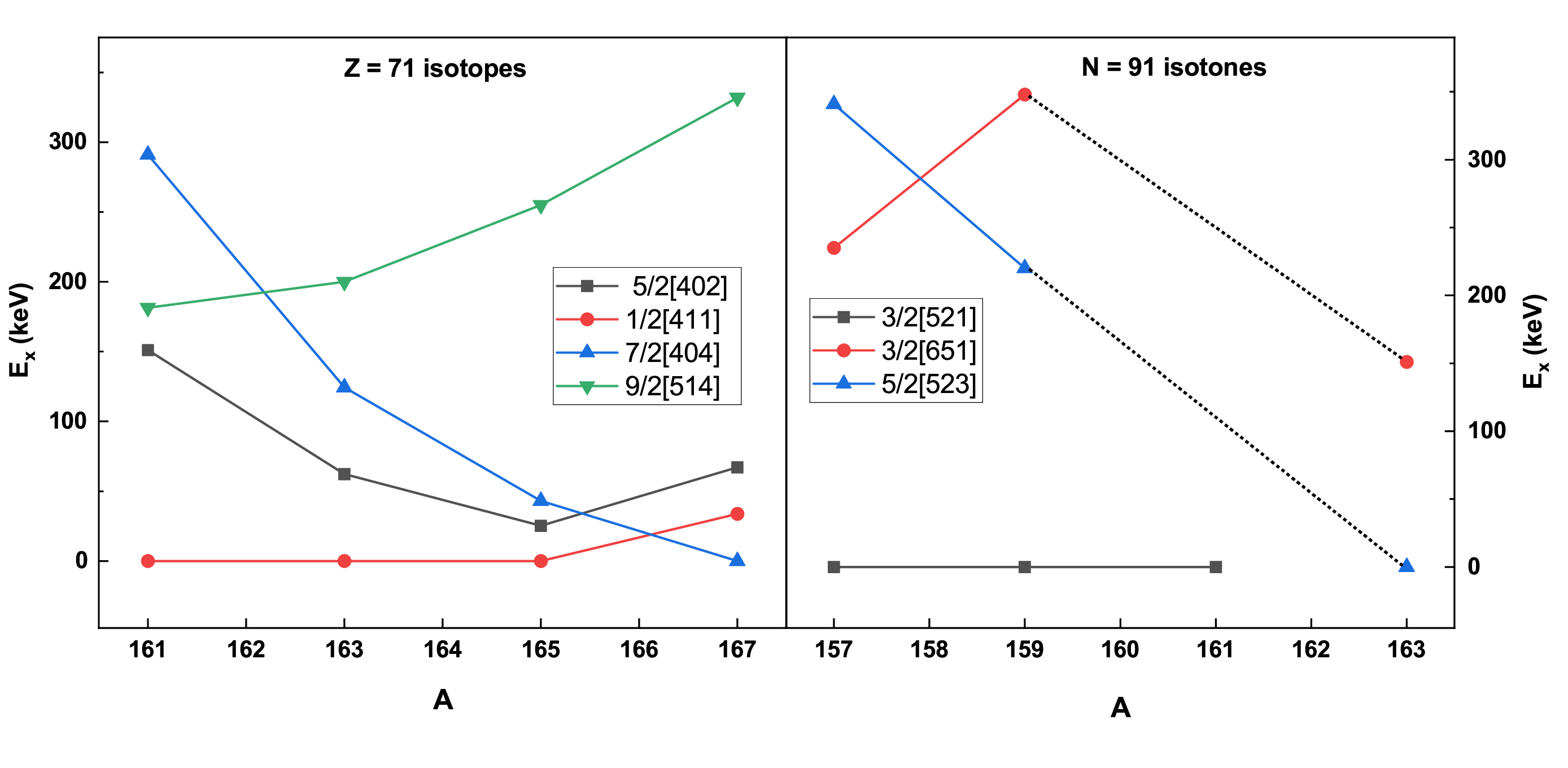}
\caption{Energy systematics of 1qp p (left) and n (right) orbitals in the mass region neighboring $^{162}$Lu. The 1qp energies were taken from the updated ENSDF datasheets \cite{ENSDF} of odd-mass Z=71 Lu isotopes and N=91 isotones in the mass region around A=162.} \label{fig1}
\end{figure*} 

Low-lying isomers also exist in neighboring odd-odd isotopes  $^{164,166,168}$Lu. The $\beta$-decay of $^{164}$Lu gs was reported by Hunter \textit{et al.} \cite{Hunt} albeit the lack of information on daughter levels populated in the decay. The parity and configuration of the gs are listed as `probable' in the ENSDF \cite{ENSDF} due to the disagreement between the theoretical magnetic moment value calculated using these assignments and the measured experimental value. Decay to the daughter nucleus levels observed and reported by Hunter \textit{et al.} \cite{Hunt} suggests the possible existence of a high-spin isomer with half-life t$_{1/2}$$\thickapprox$2 min, which could not be ascertained in the experiment. No further dedicated studies were carried out to investigate this possibility.

In contrast, the low-energy spectrum of $^{166}$Lu is well established with spin-parity and energy assignments for the gs and other low-lying levels including the isomeric states. The detailed decay study by De Boer \textit{et al.}\cite{Boer} established the existence of two short-lived isomers along with their individual $\beta$-decay branches. They assigned spin-parities and orbital configurations to these states based on the Nilsson model energy systematics and $\beta$-decay selection rules. 
In the case of $^{168}$Lu, similar assignments have been made to the gs and J$^\pi$=3$^+$ (t$_{1/2}$ = 6.7 min) isomer by Charvet \textit{et al.} \cite{Charvet}. However, the adopted level energy of this 6.7 min isomer is listed as `tentative' \cite{ENSDF} as two values were proposed based on the level's Q$_\beta$ difference with the gs \cite{Charvet} and a weak E3 transition to gs \cite{ Barci}.

It may be noted that experimental characterization of the isomeric states in  $^{162,164,168}$Lu has posed a challenge due to their short half-lives, low production cross-sections and unknown low-lying level structure. 
Model-based investigations on the level structures of these Lu isotopes can effectively describe and predict their low-lying level scheme, serving as location guides for further experimental studies. Except $^{168}$Lu, for which Barci \textit{et al.}\cite{Barci}, carried out rotor model calculations to determine the low-lying levels, every theoretical study on the said Lu isotopes have exclusively focused only on high-spin band structures and their shape deformations \cite{ShapeLu,Magicgaps,stag,164LuTMT}.

The present study aims at investigating and constructing the low-lying level scheme of $^{162}$Lu, thereby characterizing the two isomers. The favoured $\beta$-decay pathways of the gs and the two isomers to the levels of $^{162}$Yb are proposed.  We further extend our study to describe the low-lying levels and isomers in the neighboring odd-odd isotopes $^{164}$Lu, $^{166}$Lu and $^{168}$Lu and thereby assign spin-parity and configurations to some of the levels. 
To this end we employ the well-tested Two Quasiparticle Rotor Model (TQRM) \cite{Jain90,Headley98,Jain98} to determine the physically admissible 2qp bandheads in the low-energy spectra and calculate their level energies. The three-step empirical framework is discussed in Section \ref{sec2} where the energy systematics, drawn from experimental data, of the relevant single particle (1qp) proton (p) and neutron (n) orbitals in the mass region neighboring A=162 are plotted. The results from the analysis of the low-lying 2qp structure of $^{162}$Lu are described in Section \ref{sec3}.  The section that follows discusses the spin-parity, energy and configurations assignments to the gs, isomeric state and some of the low-lying levels in the odd-odd Lu isotopes $^{164}$Lu, $^{166}$Lu and $^{168}$Lu. The results of our study are summarized in Section \ref{sum}. 

\begin{table*}
\caption{\label{tab:table1}Physically admissible 2qp bandheads in $^{162}$Lu listed as GM doublets (K$_T$, K$_S$). The 1qp Nilsson orbitals of interest are listed on the top row for protons and first column for neutrons. The experimental 1qp energies of p and n orbitals as taken from neighboring nuclei are given below the orbital configuration.}
\begin{ruledtabular}
\begin{tabular}{ccccc}
 p$_i$ $\rightarrow$ &1/2$^+$[411$\downarrow$] &5/2$^+$[402$\uparrow$] &9/2$^-$[514$\uparrow$] &7/2$^+$[404$\downarrow$]\\
 n$_i$ $\downarrow$ &p$_0$=0 keV &p$_1$=62 keV &p$_2$$\thickapprox$180 keV
 &p$_3$=208 keV\\ \hline\\
 
 3/2$^-$[521$\uparrow$] &1$^-$, 2$^-$ &4$^-$, 1$^-$ &6$^+$, 3$^+$ &2$^-$, 5$^-$ \\
 n$_0$= 0 keV & & & &\\ \\
 
5/2$^-$[523$\downarrow$] &3$^-$, 2$^-$ &0$^-$, 5$^-$ &2$^+$, 7$^+$ &6$^-$, 1$^-$ \\
 n$_1$$\thickapprox$ 60 keV & & & &\\ \\

3/2$^+$[651$\uparrow$] &1$^+$, 2$^+$ &4$^+$, 1$^+$ &6$^-$, 3$^-$ &2$^+$, 5$^+$ \\
 n$_2$$\thickapprox$ 200 keV & & & &\\ 
 
\end{tabular}
\end{ruledtabular}
\end{table*}

\section{\label{sec2}Model Formulation}

 In case of deformed nuclei, the 1qp p and n states are described by the angular momentum projection quantum number  $\Omega$$^\pi$[Nn$_3$$\Lambda$$\Sigma$] where N is the principal quantum number, n$_3$ is the symmetry-axis projection of the wavefunction denoting the number of nodes, $\Lambda$ is the projection of the orbital angular momentum along the symmetry axis, and $\Sigma$ is the projection of the intrinsic spin along the symmetry axis satisfying the relation $\Omega$=$\Lambda$+$\Sigma$  \cite{BohrMottelson}. For odd-odd deformed nuclei, each 2qp configuration ($\Omega$$_p$ $\otimes$  $\Omega$$_n$) gives rise to two bands K$_\pm$ =$|$$\Omega$$_p$$\pm$$\Omega$$_n$$|$. The Gallagher–Moszkowski (GM) rule places the spins-parallel triplet  (K$_T$) state, at energy lower than the spins-anti-parallel singlet (K$_S$) state \cite{GM}. The empirical Two Quasiparticle Rotor Model (TQRM) estimates level energies in odd-odd nuclei using factors that give rise to the 2qp states. The energy of a 2qp bandhead formed by 1qp p and n orbitals $\Omega$$_p$ and $\Omega$$_n$ is deduced from the following formula \cite{Headley98,Jain98}:

\begin{equation}
    E(\Omega_p,\Omega_n)= E_0+E_p(\Omega_p)+E_n(\Omega_n)+E_{rot} +< V_{pn} >     
\end{equation}

where

E$_p$ and E$_n$ are experimental 1qp p and n orbital energies respectively, taken from the neighboring isotopes and isotones. E$_0$ is the zeroth-order two-particle energy without including the residual interaction contribution. E$_{rot}$ is the rotational energy term. The detailed formulation of this rotational energy term for an odd-odd nucleus is discussed in \cite{tqrm252Md}. V$_{pn}$ is the potential caused by the residual n-p interaction between the unpaired neutron and the proton in the odd-odd nucleus. It includes the GM splitting energy, which is the energy difference between the singlet and triplet states, and the Newby shift, which is exclusive for a K=0 state.

Essentially, the empirical TQRM calculation is a three-step process. The first step is the identification of the relevant low-lying 1qp p and  n orbitals starting from the gs E$_x$=0 keV in the neighboring odd-mass isotope/isotone. In the second step, the physically admissible 2qp bandheads arising from the coupling of the p and  n orbitals are constructed. As a final step, the bandhead energies are evaluated using Eq. (1).  A detailed description of this formalism can be found in many of our previous reports where the model has been  successfully used to describe and predict the low-lying level structures of odd-odd deformed nuclei in both the rare-earth and actinide regions \cite{tqrm154Pm,tqrm156Pm,tqrm184Ta,tqrm186Ta,tqrm240Np,tqrm244Np,tqrm252Md}.
As clarified in many of these works, the model-calculated level energies serve as ``location guides". From a broad overview of the previous works, the actual level energies were found to vary from the TQRM estimates by about 30 keV, which is considered as general uncertainty in our energy calculations. A classic example of this can be seen in our earlier work \cite{tqrm184Ta}, where the estimated energies lie within the expected energy range for the corresponding levels.
The fundamental input to the model is the experimental energies from the neighboring odd mass (A±1) isotope/isotone. The appropriate choice of the p and n states that couple to give the 2qp bandheads in an odd-odd nucleus depends on the systematics of the Nilsson orbitals in that mass region. The results of our TQRM calculations for the $^{162}$Lu isotope are discussed in the following sections. 

\section{\label{sec3}Analysis and Results}
\subsection{Low-lying level structure of $^{162}$Lu}

The systematics of the p and  n orbitals in the odd-mass neighboring Z=71 isotopes and N= 91 isotones have been carefully studied for mapping the relevant low-lying 1qp orbitals close to the Fermi surface that couple giving rise to physically admissible 2qp states in $^{162}$Lu. 
From the systematics of the 1qp p orbitals plotted in Fig.\ref{fig1}, we infer that $\pi$1/2$^+$[411$\downarrow$] consistently remains the gs in $^{161,163,165}$Lu with the first excited state being $\pi$5/2$^+$[402$\uparrow$]. The $\pi$7/2$^+$[404$\downarrow$] is observed to have a steep drop in energy as we move from A=161 to A=167. 
In view of this, the 1qp energy of this orbital is taken as the interpolated value from the trend in neighboring (A$\pm$1) nuclei. 

The available experimental data \cite{ENSDF} on the 1qp n orbital energies in the closest odd-mass isotones $^{161}$Yb and $^{163}$Hf are scarce. From Fig.\ref{fig1}, we observe that $\nu$3/2$^-$[521$\uparrow$] forms the gs in odd mass N = 91 isotones $^{157}$Dy, $^{159}$Er and $^{161}$Yb. There is a steep drop in the energy of the $\nu$5/2$^-$[523$\downarrow$] as we move from A=157 to A=163. 
In their detailed study of $^{163}$Hf rotational bands, Yadav \textit{et al.} \cite{Yadav} proposed that this n orbital drops close to the Fermi surface, becoming the gs in $^{163}$Hf. Consequently, its interpolated level energy value E$_x$$\thickapprox$60 keV is taken for our calculations. Similarly, the $\nu$3/2$^+$[651$\uparrow$] level energy is unknown at A=161. Yadav \textit{et al.} \cite{Yadav} propose this orbital energy to be around 260$\pm$50 keV in their energy plot, probably because the other levels in $^{161}$Yb above 300 keV are high-spin bands.  
From these systematics, its value is taken to be E$_x$$\thickapprox$200 keV for our calculations. These low-lying 1qp n orbital trends, independently surveyed and plotted in the present study from current ENSDF data \cite{ENSDF}, are in agreement with the results of Yadav \textit{et al.} \cite{Yadav}. 
Using these inputs, we list in Table \ref{tab:table1}, the set of physically admissible 2qp bandheads in $^{162}$Lu up to a summed E$_p$ + E$_n$ $\leq$ 400 keV.

\begin{figure*}[!t]
\centering
\includegraphics[width=14.0cm]{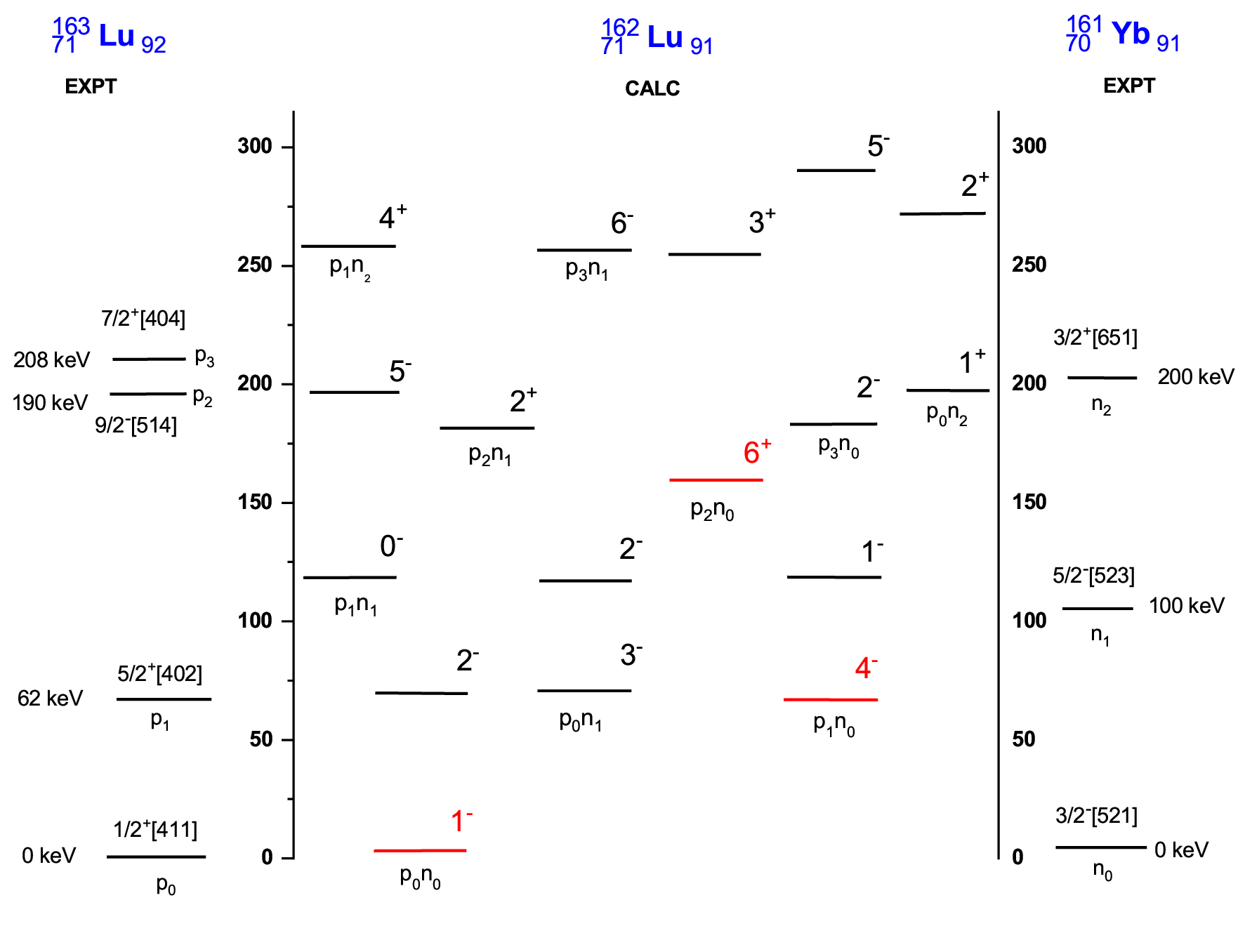}
\caption{Plot of physically admissible 2qp bandheads in $^{162}$Lu up to 300 keV calculated using TQRM. The gs and isomeric states are highlighted in red. The 1qp p and n levels giving rise to the 2qp states are placed in the left and right side respectively.} \label{fig2}
\end{figure*}

\paragraph{1.4 min gs :}Taking the inputs as described above, the low energy level scheme of $^{162}$Lu evaluated using TQRM is presented in Fig. \ref{fig2}. Half-life t$_{1/2}$$\thickapprox$ 1.37 min was adopted  for the gs of $^{162}$Lu by ENSDF evaluators \cite{ENSDF} based on multiple decay studies. The gs spin was proposed as J=1 from collinear laser spectroscopy studies reported by Georg \textit{et al.} \cite{Georg98}. The work suggested the gs parity as negative as the experimental magnetic moment matched the theoretical estimate for the coupling of 1/2$^+$[411$\downarrow$] and 3/2$^-$[521$\uparrow$] (p$_0$, n$_0$). Thus, as shown in Table \ref{tab:table1}, our TQRM analysis is in complete agreement with the adopted spin-parity for the gs, viz., 
\\

\textbf{$^{162}$Lu$^{gs}$:} 1$^-$\{$\pi$1/2$^+$[411$\downarrow$]$\otimes$$\nu$3/2$^-$[521$\uparrow$]\}   
\\

\paragraph{1.5 min isomer}: The current data sheet lists the first isomeric state with a tentative J$^\pi$ = (4$^-$) and no energy value. The ENSDF evaluators suggest the possible spin and configuration of this state ``by analogy with the situation in $^{164}$Lu"\cite{ENSDF} wherein, similar pair of states (1$^-$ and 4$^-$) is observed. As seen in Fig. \ref{fig2}, the p$_1$$\otimes$n$_0$: J$^\pi$ = 4$^-$ level is determined to be just above the gs at E$_x$$\thickapprox$ 62 keV. With $\Delta$I and $\Delta$K=3, a $\gamma$ transition from this level to gs will be highly hindered, rendering an isomeric character to this state. The J$^\pi$=2$^-$ singlet partner of the gs GM doublet and the p$_0$$\otimes$n$_1$: J$^\pi$ = 3$^-$ level are expected to lie slightly above this level. Hence, we confirm the J$^\pi$ and configuration for the 1.5 min isomer of $^{162}$Lu and propose its energy as:
\\

\textbf{$^{162}$Lu$^{m1}$:}4$^-$\{$\pi$5/2$^+$[402$\uparrow$]$\otimes$$\nu$3/2$^-$[521$\uparrow$]\}; \textbf{E$_x$}$\thickapprox$62 keV.   
\\

\paragraph{1.9 min isomer:} The only study to date that reports isomeric states in $^{162}$Lu is the decay spectroscopy by Behrens \cite{Behrens} wherein, in addition to the 1.5 min isomer, a 1.9 min activity in the $^{162}$Lu $\epsilon$+$\beta^+$-decay was reported. While the J$^\pi$=4$^-$ state, confirmed as the 1.5 min isomer, is akin to the neighboring odd-odd isotope $^{164}$Lu, the second isomer in $^{162}$Lu is an entirely new level to be identified, as it is listed in the current data sheets with no J$^\pi$ or energy assignments \cite{ENSDF}. From the model constructed level scheme (Fig. \ref{fig2}), we propose the p$_3$n$_0$: J$^\pi$= 6$^+$ 2qp bandhead as the probable candidate for this isomer, based on the following inferences:

i) This J$^\pi$= 6$^+$ state does not have any level below it with a spin difference of $\Delta$K$\leq$1 for an unhindered decay. 

ii) The J$^\pi$=6$^+$ is the level with highest spin at energies  E$_x$$\leq$200 keV. There are no other states with higher spins close to it. We may also note from Fig. \ref{fig1} that the $\pi$9/2$^-$[514$\uparrow$] orbital forming this state appears at lower energies only in $^{161}$Lu and $^{163}$Lu, beyond which it is found at energies above 250 keV. There are no other intruder 1qp p or  n orbitals at such low energies which can couple and form high-spin 2qp states. 

iii) An M2(+E3) transition to the J$^\pi$=4$^-$ (1.5 min isomer) or an E1 transition to J$^\pi$K=5$^-$4 can be expected from this isomeric state, albeit with $\Delta$K=2. However, according to Alaga's $\gamma$-decay selection rules for deformed nuclei \cite{Alaga1957}, this J$^\pi$=6$^+$$\rightarrow$ 4$^-$ (M2) IT decay with 1qp orbital transformation $\pi$9/2$^-$[514$\uparrow$]$\rightarrow$$\pi$5/2$^+$[402$\uparrow$] would be a hindered decay. Hence, the J$^\pi$=6$^+$ state can be expected to have a longer half-life. 

Under these considerations, we propose the following assignment to the 1.9 min isomer:

\begin{center}
\textbf{$^{162}$Lu$^{m2}$:} 6$^+$\{$\pi$9/2$^-$[514$\uparrow$]$\otimes$$\nu$3/2$^-$[521$\uparrow$]\}; 

\textbf{E$_x$}$\thickapprox$160 keV.
    
\end{center}

It may be worthwhile to point out that, in a broad survey of M2 IT decays \cite{Atlas}, isomer half-lives were observed to range from ns to s, except for nuclei with high K hindrances. Considering that the half-life assignment t$_{1/2}$ = 1.9 min was reported in the lone experimental study \cite{Behrens}, where no IT decay was observed, this assignment would require further experimental validation. 

\paragraph{Other levels:} Few other levels determined in our TQRM calculations are also in agreement with the current adopted low-lying level scheme of $^{162}$Lu \cite{ENSDF}. Hild \textit{et al.}\cite{Hild162} proposed a level at E$_x$= 196 keV which was assigned tentative J$^\pi$ = (1$^+$) based on the observation of unhindered $\beta$-decay from the parent nucleus $^{162}$Hf with log \textit{ft} $\thickapprox$ 4.1. As seen in Fig. \ref{fig2}, the p$_0$$\otimes$n$_2$: J$^\pi$=1$^+$ \{$\pi$1/2$^+$[411$\downarrow$]$\otimes$$\nu$3/2$^+$[651$\uparrow$]\} level at E$_x$$\thickapprox$199 keV would be the likely candidate to which an allowed $\beta$-decay ($\Delta$J=1, $\Delta$$\pi$=no) from the parent $^{162}$Hf gs (J$^\pi$=0$^+$) is possible. We hence assign this spin-parity and configuration to the reported 196 keV level in $^{162}$Lu.
Our calculations do not ascertain the existence of a level at E$_x$$\thickapprox$5 keV as proposed by Hild \textit{et al.} \cite{Hild162}. The non-existence of the E$_x$=5 keV level could possibly require a rearrangement of the 191 keV $\gamma$-transition within the existing level scheme. In the updated data sheets \cite{NDS,ENSDF}, a level with E$_x$$\thickapprox$ 79 keV is placed, based on a reported $\gamma$-decay to the gs with the same energy. The p$_0$$\otimes$n$_1$ J$^\pi$= 3$^-$\{$\pi$1/2$^+$[411$\downarrow$$\otimes$5/2$^-$[523$\downarrow$]\} is a possible candidate for this level as it can decay to the gs through an E2 transition. A more detailed experimental study is warranted to unambiguously establish the low-lying structure of $^{162}$Lu for which our model calculations can serve as location guides.  

\subsection{\label{sec4}Favored decay branches of gs and isomers}

The fundamental $\beta$-decay selection rules defines the ‘forbiddenness’ of any decay based on the spin-parity of the parent and daughter levels as:

\begin{center}
 Allowed:    	$\Delta$I=0,1 and $\Delta$$\pi$=no  

First forbidden:  	$\Delta$I=0,1 and $\Delta$$\pi$=yes   
\end{center}

Alaga \cite{Alaga1957} extended these rules to deformed nuclei whose 1qp p and n levels are defined by Nilsson orbital quantum numbers $\Omega$$^\pi$[Nn$_3$$\Lambda$$\Sigma$]. Based on comparisons of multiple $\beta$-decays with different particle configurations, the said work proposed the following selection rule, defining ‘allowed unhindered’ (au) decays in deformed nuclei:
\begin{center}
   au: $\Delta$$\Omega$= 0,1 and $\Delta$N = $\Delta$n$_3$ =$\Delta$$\Lambda$ = 0  
\end{center}

These `au' decays were found to have log \textit{ft} values ranging from 4.5-5.0 according to available experimental data. Later studies extended $\beta$-decay selection rules for deformed nuclei, based on their observation of transitions with lesser experimental log \textit{ft}, despite not following the Alaga rule.
In the actinide and rare-earth regions, Sood \& Sheline \cite{SS89sf} identified the existence of `spin-flip' $\beta$-decays in the 147$\leq$A$\leq$190 mass region  where orbitals with unchanged asymptotic quantum numbers N=n$_3$+$\Lambda$=5 but opposite spins were observed to undergo allowed decays (log $ft\leq$5.2). In a later work, Sood \& Sheline \cite{SS92ext} extended the selection rules  ($\Delta$n$_3$$<$2) for actinides citing 30 decays which were observed to have log $ft<$6.2 and hence were less hindered. An additional selection rule was defined as $\Delta$N=0, $\Delta$n$_3$+$\Delta$$\Lambda$=0 and $\mid$$\Delta$n$_3$$\mid$$\leq$2 when 1qp p and n orbitals were found to have asymptotic quantum numbers N=n$_3$+$\Lambda$=6. 

Further, in a preliminary report of $\beta$-decays in the medium heavy rare earth region, Gowrishankar \textit{et al.}\cite{DAE1ff} observed the existence of decays with log $ft\leq$6.3 despite violating the Alaga selection rule. From the extensive compilation of $\beta$-decay data by Singh \textit{et al.} \cite{singh98}, in the mass region A=157-167, around 20 such cases were identified. This preliminary observation led to the definition of a new possible category, viz., the first forbidden fast transitions (1ff) with the following selection rules:
\begin{center}
    1ff: $\Delta$$\pi$=yes and	 $\Delta$N=($\Delta$n$_3$+$\Delta$$\Lambda$) = $\pm$1

Allowed fast: $\Delta$N=($\Delta$n$_3$+$\Delta$$\Lambda$) = 0
\end{center}

The log \textit{ft} values of these 1ff transitions were found to be lower than that of first forbidden transitions, indicating a lesser hindered decay branch. The decay of gs and isomers in deformed nuclei are generally governed by the selection rules of Alaga \cite{Alaga1957} but exceptional cases of allowed fast and 1ff transitions are intriguing anomalies that need rigorous study and characterization.

\begin{figure}[h]
\centering
\includegraphics[width=1\columnwidth]{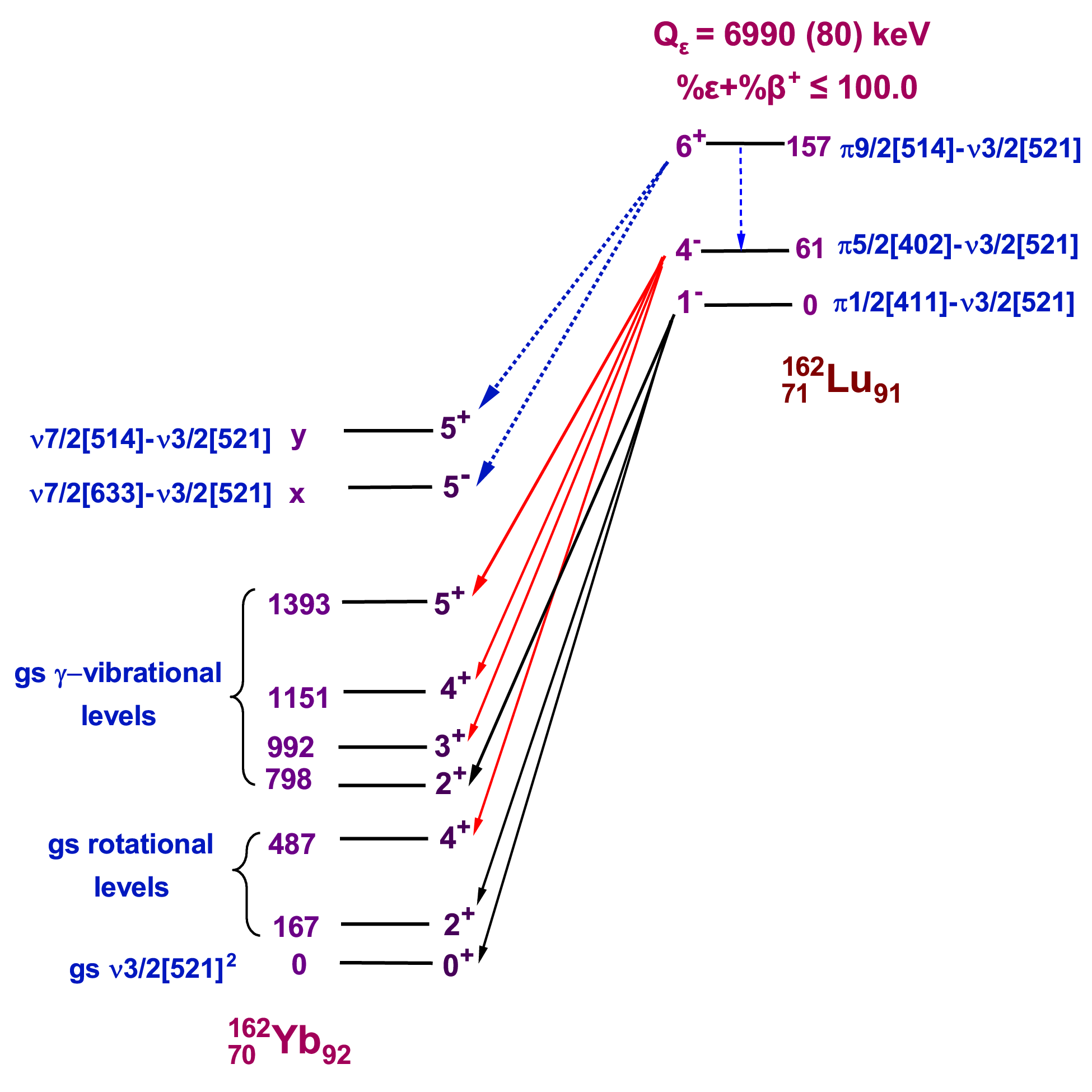}
\caption{Favored $\beta$-decay branches of the $^{162}$Lu gs and isomeric states to some of the daughter levels (not to scale) in $^{162}$Yb based on the $\beta$-decay selection rules for spin-parity and orbital configuration assignments made in the present work. The Q$_\beta$ and percentage $\beta$ decay data have been taken from ENSDF \cite{ENSDF}.} \label{fig3}
\end{figure}

The available data on the $\beta$-decay of $^{162}$Lu to the daughter states of $^{162}$Yb are based on a single study carried out by Behrens \cite{Behrens}, which was later elaborated by McCutchen \textit{et al.} \cite{McCutchen}. They collectively reported the daughter levels in $^{162}$Yb populated in the decay without log \textit{ft} and decay intensities. 
As adopted in the latest data sheets \cite{ENSDF,NDS}, the gs and isomeric states of $^{162}$Lu are reported to undergo independent $\beta$-decays ($\beta^+ + \epsilon$), while their explicit branching is unknown. 
Over the last two decades, no further attempts have been made to delineate the $\beta$-decays of the $^{162}$Lu gs and isomers. Using the spin-parity and orbital configurations assigned in our present study, we herewith propose the possible $\beta$-decay branches for the gs and isomeric states of $^{162}$Lu to the adopted levels \cite{ENSDF} of $^{162}$Yb. While we restrict our analysis to the levels in $^{162}$Yb for which the configurations is known, decay to higher energy levels are also possible as the Q-value of the parent decay is high. 
The identification of favored $\beta$-decay branches from each long-lived state are discussed in detail.  

\begin{figure*}[!t]
\centering
\includegraphics[width=14.0cm]{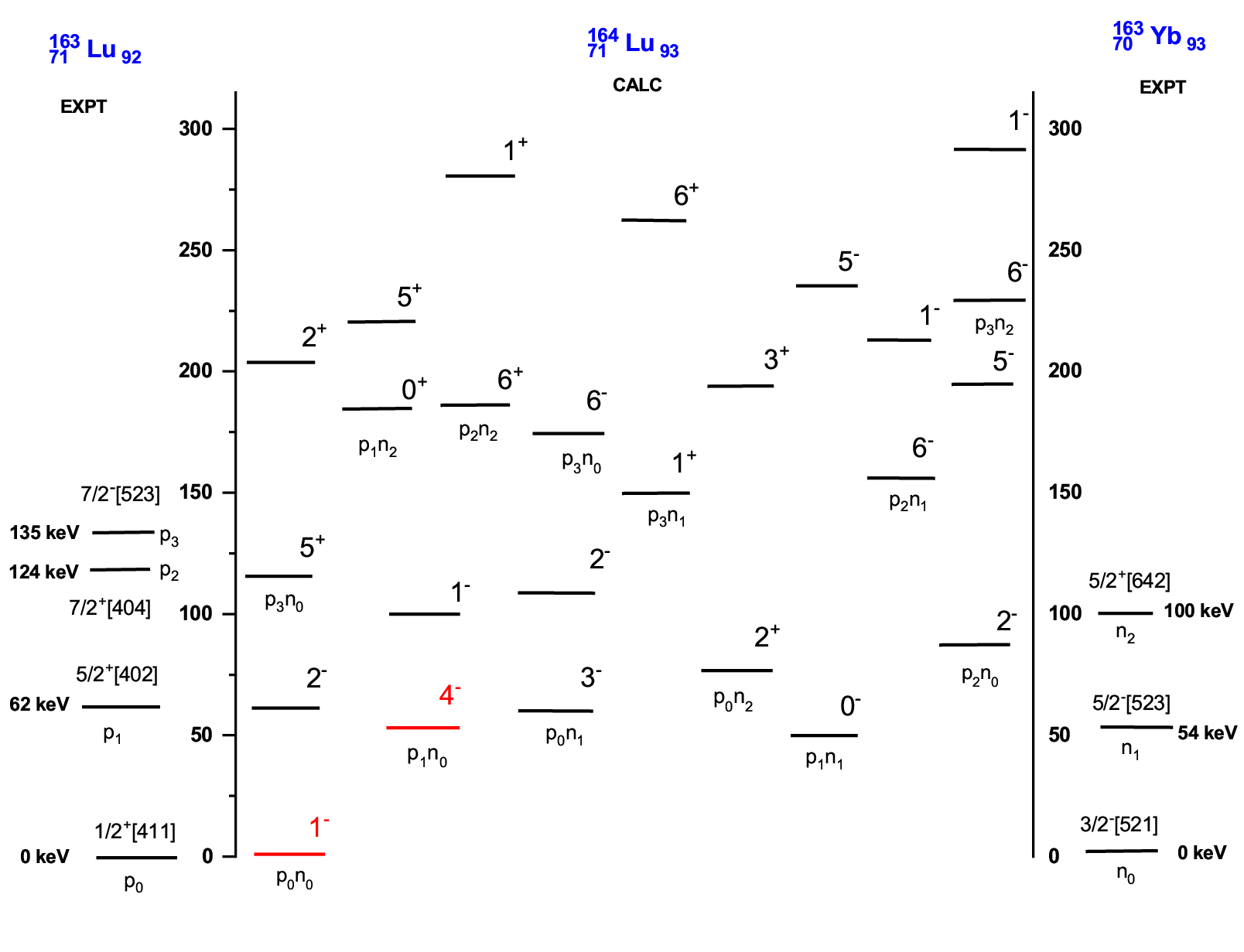}
\caption{Plot of model calculated energies of 2qp bandheads in $^{164}$Lu (center) upto E$_x$=300 keV along with the 1qp p and  n orbital energies taken from isotopic neighbor $^{163}$Lu (left) and isotonic neighbor $^{163}$Yb (right). The gs and possible isomeric state as proposed in previous decay study \cite{Hunt} are highlighted (in red).} \label{fig4}
\end{figure*}

\paragraph{Decay of 1.37 min $^{162}$Lu$^{gs}$}: With J$^\pi$=1$^-$ assignment, the gs $\beta$- decay will be favored to daughter levels having spins J= 0, 1, 2 through a p$\rightarrow$n transformation. 
The $\nu$3/2$^-$[521$\uparrow$] is the 1qp valence neutron state, as observed from the energy systematics in the neighboring mass region. Thereby, the gs of $^{162}$Yb has the 2-neutron configuration $\nu$3/2$^-$[521$\uparrow$]$^2$. Accordingly, the 1.37 min gs of $^{162}$Lu can  undergo `1ff' $\beta$-decays to the gs J$^\pi$=0$^+$, J$^\pi$K=2$^+$0 and J$^\pi$=2$^+$ gs $\gamma$ vibrational bandhead of the daughter nucleus. 

\paragraph{Decay of the 1.5 min isomer:} The favored $\beta$-decay of the 1.5 min isomer can populate $^{162}$Yb levels with spins J= 3, 4 and 5. Our configuration assignment of J$^{\pi}$=4$^-$\{$\pi$5/2$^+$[402$\uparrow$]$\otimes$$\nu$3/2$^-$[521$\uparrow$]\} to this isomer warrants a p$\rightarrow$n transformation $\pi$5/2$^+$[402$\uparrow$]$\rightarrow$ $\nu$3/2$^-$[521$\uparrow$] resulting in a `1ff' decay as listed in Gowrishankar \textit{et al.} \cite{DAE1ff}. Hence we propose the $\beta$-branching to the J$^\pi$=4$^+$ (487 keV) excited state of the gs rotational band and the J$^\pi$=3$^+$, 4$^+$, 5$^+$ levels of the vibrational band in $^{162}$Yb. 

\paragraph{Decay of the 1.9 min isomer:} 
There are two possible modes by which the J$^\pi$=6$^+$ isomer can undergo decay. As discussed previously, an IT decay to the J$^\pi$=4$^-$ state is possible, resulting in a E$_\gamma$$\thickapprox$100 keV hindered M2 transition. 
The other mode is a direct $\beta$-decay to the daughter levels of $^{162}$Yb. Alaga's selection rule does not favor $\beta$-decay from the $\pi$9/2$^-$[514$\uparrow$]$\otimes$$\nu$3/2$^-$[521$\uparrow$] configuration to any level in the $^{162}$Yb gs band due to the large spin difference in a $\pi$9/2$^-$$\rightarrow$$\nu$3/2$^-$ transformation.
A survey of the Nilsson 1qp n and p orbitals near N=92 and Z=70 region respectively, shows $\nu$5/2$^-$[523$\downarrow$], $\nu$1/2$^+$[660$\uparrow$], $\nu$3/2$^+$[651$\uparrow$], $\nu$5/2$^+$[642$\uparrow$] close to the Fermi surface, with two high-spin orbitals, $\nu$7/2$^-$[514$\downarrow$] and $\nu$7/2$^+$[633$\uparrow$] at higher energies. The 1.9 min isomer can undergo $\beta$-decay to high-spin states with the following 2-n configuration:

\begin{center}
$\nu$7/2$^-$[514$\downarrow$]$\otimes$$\nu$3/2$^-$[521$\uparrow$] \\
$\nu$7/2$^+$[633$\uparrow$]$\otimes$$\nu$3/2$^-$[521$\uparrow$]
\end{center}

The first p$\rightarrow$n transition, $\pi$9/2$^-$[514$\uparrow$]$\rightarrow$$\nu$7/2$^-$[514$\downarrow$] would be an allowed spin-flip decay which could have a log $ft<$ 5.2 \cite{SS89sf}. 
On the other hand, $\pi$9/2$^-$[514]$\rightarrow$7/2$^+$[633$\uparrow$] is a 1ff transition with change in parity and $\Delta$N=$\Delta$n$_3$+$\Delta$$\Lambda$= -1. 
It may also be noted that these two daughter levels have not been experimentally reported so far in $^{162}$Yb, probably as they are expected at very high energies. Nevertheless, the proposed $\beta$-decay is consistent with the high Q$_\beta$ value \cite{ENSDF} of $^{162}$Lu.   
The proposed $\beta$-decay branches of the $^{162}$Lu gs and isomers to select levels of daughter nucleus $^{162}$Yb is illustrated in Fig. \ref{fig3}.

\section{Isomers in neighboring odd-odd isotopes}

From the available experimental data. it is worth noting that around $^{162}$Lu there is a change in the mode of decay among the Lu isotopes. Close to the proton drip line, the lighter-mass Lu isotopes compete to decay through $\alpha$- or $\beta$-decay. Starting from A=161, the Lu isotopes progressively exhibit only $\epsilon$+$\beta^+$ decay all the way up to their stable isotope $^{175}$Lu beyond which neutron-rich isotopes tend to undergo $\beta^-$-decay. Among the purely $\epsilon$+$\beta^+$-decaying odd-odd Lu isotopes, $^{170}$Lu was studied using Particle Rotor Model calculations by Sood \textit{et al.} \cite{Sood170Lu}. The work characterized the low-lying level structure, including a reported isomer, and also predicted the possible existence of a new low-lying high-spin isomer. Among the $\beta^+$-decaying Lu isotopes, the experimental low-energy spectrum is extensively reported for A$\geq$170 albeit with some tentative assignments. Whereas the gs and two isomers in $^{166}$Lu have been unambiguously characterized and their assignments firmly established, the low-lying levels of $^{164,168}$Lu remain without definitive energy, J$^\pi$ and configuration assignments. For example, in $^{164}$Lu Hunter \textit{et al.} \cite{Hunt} proposed a possible isomeric state which is not adopted in the datasheets due to lack of proper experimental evidence. Similarly, in $^{168}$Lu, the energy of the 6.7 min isomer is tentative. As an extension of our investigations in $^{162}$Lu, we attempt to characterize the low-lying isomers reported in these neighboring odd-odd nuclei. Existing data, ambiguities, our model calculations and the results therein are discussed in the following subsections.   

\subsection{$^{164}$Lu} 

The initial report on the production and decay of $^{164}$Lu by Neiman and Ward \cite{NW} determined the gs half-life to be t$_{1/2}$= 4.1 min. The established low-lying level structure of $^{164}$Lu is primarily based on the decay study of its parent nucleus $^{164}$Hf conducted by Hild \textit{et al.} \cite{Hild164}. Further, Hunter \textit{et al.} \cite{Hunt} investigated the decay of $^{164}$Lu to the daughter states of $^{164}$Yb. Their analysis suggested that the gs of $^{164}$Lu is formed by the 1qp  n orbital $\nu$3/2$^-$[521$\uparrow$] inferred from the absence of fast $\beta$-transitions to the daughter levels. The study also identified the existence of another shorter half-life component,  indicating the possible existence of an isomer.  

\par The evaluated data sheets \cite{ENSDF} list the gs spin and probable orbital configuration based on laser spectroscopy studies by Georg \textit{et al.}\cite{Georg98}, but assign a tentative negative parity due to disagreement between theoretical magnetic moment for the proposed configuration and the experimental value. Our TQRM calculations confirm the proposed orbital configuration of $^{164}$Lu gs to be $\pi$1/2$^+$[411$\downarrow$]$\otimes$$\nu$3/2$^-$[521$\uparrow$]. The involvement of the $\nu$3/2$^-$[521$\uparrow$] orbital in the gs, proposed by Hunter \textit{et al.} \cite{Hunt}, is corroborated in our analysis of 1qp p and n orbital systematics. Using our procedure of identifying the applicable 1qp p and n orbitals, the energies of the physically admissible 2qp bandheads are plotted in Fig. \ref{fig4}. 

In their study, Hunter \textit{et al.} reported a weak population of some high-spin states in $^{164}$Yb with a comparatively shorter half-life. Some of the $^{164}$Yb $\gamma$-rays were found to have a half-life of around 2 min, which the work mentions as ``evidence for the existence of a high-spin isomer". However, this activity was reported to be very weak in the experiment for further characterization. As seen in our level scheme (Fig. \ref{fig4}), the J$^\pi$=4$^-$ (p$_1$n$_0$) state lies close to the gs at E$_x$$\thickapprox$52 keV. Its decay to either the gs or the J$^\pi$=0$^-$ level slightly below it would require an E3 or M4 $\gamma$ transition, respectively. This state could well be the 2 min isomer, that was proposed by Hunter \textit{et al.} \cite{Hunt}.

Although the latest data sheets \cite{ENSDF} list only five levels below 300 keV, theoretical considerations and empirical calculations reveal that up to 24 2qp bandheads can be expected in $^{164}$Lu within this energy range, as illustrated in Fig 4. Based on the observed $\gamma$ transitions to the gs, the data sheets list two levels, one at 31 keV and another at 153 keV with spin J$\leq$ 3. These could well be the rotational levels of the 2qp bandheads (p$_0$n$_0$) and (p$_1$n$_0$) respectively, the former having an anology in $^{162}$Tm. Experimental studies aimed at unraveling the low-lying level structure of $^{164}$Lu would be of great importance to confirm the existence of the proposed isomer and characterize the unknown rotational levels. 

\subsection{$^{166}$Lu} The low-lying level structure of $^{166}$Lu was reported by De Boer \textit{et al.} \cite{Boer}  delineating the J$^\pi$, level energies, orbital configurations and individual $\beta$-decay branches of the gs and two isomers. The gs of this nucleus is J$^\pi$= 6$^-$\{$\pi$1/2$^+$[411$\downarrow$]$\otimes$$\nu$3/2$^-$[521$\uparrow$]\} with  t$_{1/2}$=2.65 min. Two low-lying isomers are listed in the current ENSDF \cite{ENSDF} with t$_{1/2}$=1.41 min and t$_{1/2}$= 2.12 min. While their spins were established through collinear spectroscopy studies \cite{Georg98}, the parity of the 1.41 min isomer is reported tentatively due to the unconfirmed (M3) multipolarity of the IT decay to the gs. Our model calculations place the  J$^\pi$= 3$^-$\{$\pi$1/2$^+$[411$\downarrow$]$\otimes$$\nu$5/2$^-$[523$\downarrow$]\} level at E$_x$$\thickapprox$ 39 keV, in agreement with the listed \cite{ENSDF} configuration and energy. Hence we confirm the negative parity assignment to the 1.41 min isomer in $^{166}$Lu. Similarly, our calculations place the J$^\pi$= 0$^-$\{$\pi$5/2$^+$[402$\uparrow$]$\otimes$$\nu$5/2$^-$[523$\downarrow$]\} level at energy E$_x$$\thickapprox$ 41 keV, close to the experimentally reported energy of the 2.12 min isomer. 

In addition to the isomers and high spin rotational levels, the adopted data of $^{166}$Lu lists five levels below E$_x$=150 keV, placed on the basis of decay $\gamma$-rays \cite{Boer}. Some of the physically admissible 2qp bandheads calculated using TQRM were found to be in concurrence with these listed levels. 
For example, our calculations estimate the energy of the J$^\pi$=1$^+$\{$\pi$7/2$^-$[523$\uparrow$]$\otimes$5/2$^-$[523$\downarrow$]\} state as E$_x$$\thickapprox$128 keV which corroborates with the adopted energy and configuration assignment.  
Similarly, a level with E$_x$= 57.2 keV reported with a tentative spin assignment J$^\pi$=(1)$^-$, is identified as the singlet partner J$^\pi$=1$^-$\{$\pi$7/2$^+$[404$\downarrow$]$\otimes$ $\nu$5/2$^-$[523$\downarrow$]\} of the gs. Our energy estimate of this level is E$_x$=52 keV, confirming its J$^\pi$ and configuration. Due to the sparse availability of experimental 1qp energies in the A$\pm$1 neighboring nuclei, our model calculations are limited to only these levels.  However, the close agreement between the available experimental data \cite{ENSDF} and our calculated values for these $^{166}$Lu levels substantiates the validity of the model.    

\subsection{$^{168}$Lu}As seen in the updated data sheets \cite{ENSDF}, the assignments to most of the low-lying energy levels in $^{168}$Lu remain ambiguous, except for spin assignments to the gs and the only known 6.7 min isomeric state. The gs of $^{168}$Lu is listed as J$^\pi$=6$^{(-)}$\{$\pi$7/2$^+$[404$\downarrow$]$\otimes$$\nu$5/2$^-$[523$\downarrow$\} \cite{ENSDF}. The gs spin has been confirmed by collinear spectroscopy \cite{Georg98} and its orbital configuration is proposed based on the measured magnetic moment of the state. The observed 1qp p and n orbital energy systematics in neighboring isotopes and isotones agree with this orbital configuration, and hence we confirm the negative parity assignment for the gs. 

The 6.7 min isomeric state of $^{168}$Lu was first discovered and placed at a tentative energy E$_x$$\thickapprox$240$\pm$140 keV by Charvet \textit{et al.} \cite{Charvet} through $\beta$-$\gamma$ coincidence observations. The energy was proposed based on the difference in Q$_\beta$ values of the gs and the isomeric state. Its spin and parity were proposed by Ekstrom \textit{et al.} \cite{Ekstrom} as J$^\pi$= 3$^+$ through atomic beam studies and supported by the decay studies of Charvet \textit{et al.} \cite{Charvet}. 
In a later $\gamma$-decay study of $^{168}$Lu, Barci \textit{et al.} \cite{Barci} reported the observation of a weak E$_{\gamma}$$\thickapprox$ 202 keV transition, proposing it as isomeric transition despite terming it as `doubtful'.  
Due to this uncertainty, the authors contradicted the existing assignment by assuming the J$^\pi$=3$^+$ state as gs and proposed its energy relative to  J$^\pi$=6$^-$ state as E$_x$$\thickapprox$ 202 keV. In the later part of their study, Barci \textit{et al.} \cite{Barci} also performed particle rotor model calculations identifying 72 possible $^{168}$Lu states. Their calculations proposed the relative energy of the J$^\pi$=3$^+$ state to be E$_x$$\thickapprox$110(50) keV, substantially differing from their own experimental proposition. 
Our model calculated level energy for this isomer with J$^\pi$=3$^+$\{$\pi$1/2$^-$[541$\downarrow$]$\otimes$$\nu$5/2$^-$[523$\downarrow$\}, turns out to be E$_x$ $\thickapprox$100 keV, closely agreeing with the calculations of Barci \textit{et al} \cite{Barci}. Another observation of interest in this context is the violation of GM rule in the bandheads formed by $\nu$5/2$^+$[642$\uparrow$]. Detailed experimental studies of high-spin band structures in $^{168}$Lu \cite{Ha1998,Ha2002,Roux2021} reported that the $\pi$g$_{7/2}$$\otimes$$\nu$i$_{13/2}$ bandheads in $^{168}$Lu and its neighboring nuclei had high-spin K$_S$ state at lower energy relative to the low-spin K$_T$ state in a GM doublet. The absence of favorable bandheads at lower energy justifies the isomeric nature of the J$^\pi$=3$^+$ level at E$_x$$\thickapprox$100 keV. 

\par The J$^\pi$, orbital configuration, and energy assignments of the gs and low-lying isomers in $^{162,164,166,168}$Lu are listed in Table \ref{tab:table3}. From the systematics of the n and p orbitals, we see a consistent pattern in the formation of these isomers. In the mass region A=160-170, the low-lying 1qp p orbitals $\pi$1/2$^+$[411$\downarrow$], $\pi$5/2$^+$[402$\uparrow$] and $\pi$7/2$^+$[404$\downarrow$] are closely spaced. Their energy difference ($\Delta$E$_p$) varies within 20-60 keV. The gs and isomers in $^{162,164}$Lu and in $^{166,168}$Lu are formed majorly by the coupling of 1qp  n orbital $\nu$3/2$^-$[521$\uparrow$] and $\nu$5/2$^-$[523$\downarrow$] with the above p orbitals respectively. 

\begin{table}
\centering
\caption{\label{tab:table3} Low-lying isomers in the odd-odd Lu isotopes of interest along with their their spin-parities, orbital configuration and energies as obtained in the present study using TQRM.}
\begin{ruledtabular}
\begin{tabular}{cccc}
Nucleus &J$^\pi$ &Orbital &Calculated \\ 
& & configuration &Energy (keV) \\
\hline\\
 $^{162}$Lu &1$^-$ &$\pi$1/2$^+$[411$\downarrow$]$\otimes$$\nu$3/2$^-$[521$\uparrow$] &0\\
  &4$^-$ &$\pi$5/2$^+$[402$\uparrow$]$\otimes$$\nu$3/2$^-$[521$\uparrow$] &62\\
  &6$^+$ &$\pi$9/2$^-$[514$\uparrow$]$\otimes$$\nu$3/2$^-$[521$\uparrow$] &157\\
  \\
$^{164}$Lu &1$^-$ &$\pi$1/2$^+$[411$\downarrow$]$\otimes$$\nu$3/2$^-$[521$\uparrow$] &0\\
 &4$^-$ &$\pi$5/2$^+$[402$\uparrow$]$\otimes$$\nu$3/2$^-$[521$\uparrow$] &52$^*$\\
\\
$^{166}$Lu &6$^-$ &$\pi$7/2$^+$[404$\downarrow$]$\otimes$$\nu$5/2$^-$[523$\downarrow$] &0\\
 &3$^-$ &$\pi$1/2$^+$[411$\downarrow$]$\otimes$$\nu$5/2$^-$[523$\downarrow$] &39\\
 &0$^-$ &$\pi$5/2$^+$[402$\uparrow$]$\otimes$$\nu$5/2$^-$[523$\downarrow$] &41\\
\\
$^{168}$Lu &6$^-$ &$\pi$7/2$^+$[404$\downarrow$]$\otimes$$\nu$5/2$^-$[523$\downarrow$] &0\\
 &3$^+$ &$\pi$1/2$^-$[541$\downarrow$]$\otimes$$\nu$5/2$^-$[523$\downarrow$] &100\\ \\
\end{tabular}
\end{ruledtabular}
$^*$ Possible isomer proposed by Hunter \textit{et al.} \cite{Hunt}
\end{table}

\section{\label{sum}Summary}
The low-lying level structure of the odd-odd nucleus $^{162}$Lu was investigated using the Two Quasiparticle Rotor Model (TQRM). The spin-parity and orbital configurations of the gs and isomers have been confirmed, and the level energies of the isomers have
been assigned for the first time.  
The 1.4 min isomer was confirmed to have J$^\pi$= 4$^-$\{$\pi$5/2$^+$[402$\uparrow$]$\otimes$$\nu$3/2$^-$[521$\uparrow$]\} at E$_x$$\thickapprox$ 62 keV, and the 1.9 min isomer has been assigned J$^\pi$= 6$^+$\{$\pi$9/2$^-$[514$\uparrow$]$\otimes$$\nu$3/2$^-$[521$\uparrow$]\} at E$_x$$\thickapprox$157 keV. Individual $\beta$-decay branches for the gs and isomers, previously unknown, were proposed based on our assignments and the $\beta$-decay selection rules for deformed nuclei. 
\par The analysis was extended to neighboring odd-odd Lu isotopes, to address ambiguities in their low-lying levels. For $^{164}$Lu, a possible J$^\pi$= 4$^-$\{$\pi$5/2$^+$[402$\uparrow$]$\otimes$$\nu$3/2$^-$[521$\uparrow$]\} isomer at E$_x$$\thickapprox$52 keV was suggested, in agreement with earlier tentative observations. For $^{168}$Lu, the energy of the 6.7 min isomer, previously listed with large uncertainty, was refined to E$_x$$\thickapprox$100 keV. The occurrence of multiple low-lying isomers in these isotopes is attributed to the low-energy trend of the corresponding 1qp p orbitals in the A=160-170 mass region.The assignments for $^{162}$Lu and its even-mass isotopic neighbors from the present study would serve as location guides for further $\beta$-decays and low-energy spectroscopic studies, aiding the systematic understanding of isomer formation in deformed odd-odd nuclei and can be substantiated by microscopic model-based studies using DFT or Mean-field calculations . 

\textbf{Acknowledgments:} The authors dedicate this work to their revered founder Chancellor, Bhagawan Sri Sathya Sai Baba and gratefully acknowledge the inputs and guidance of Prof P. C. Sood. NS acknowledges the fellowship from UGC-SJSGC scheme. 

\nocite{*}

\bibliography{apssamp}

\begin{thebibliography}{52}%
\makeatletter
\providecommand \@ifxundefined [1]{%
 \@ifx{#1\undefined}
}%
\providecommand \@ifnum [1]{%
 \ifnum #1\expandafter \@firstoftwo
 \else \expandafter \@secondoftwo
 \fi
}%
\providecommand \@ifx [1]{%
 \ifx #1\expandafter \@firstoftwo
 \else \expandafter \@secondoftwo
 \fi
}%
\providecommand \natexlab [1]{#1}%
\providecommand \enquote  [1]{``#1''}%
\providecommand \bibnamefont  [1]{#1}%
\providecommand \bibfnamefont [1]{#1}%
\providecommand \citenamefont [1]{#1}%
\providecommand \href@noop [0]{\@secondoftwo}%
\providecommand \href [0]{\begingroup \@sanitize@url \@href}%
\providecommand \@href[1]{\@@startlink{#1}\@@href}%
\providecommand \@@href[1]{\endgroup#1\@@endlink}%
\providecommand \@sanitize@url [0]{\catcode `\\12\catcode `\$12\catcode `\&12\catcode `\#12\catcode `\^12\catcode `\_12\catcode `\%12\relax}%
\providecommand \@@startlink[1]{}%
\providecommand \@@endlink[0]{}%
\providecommand \url  [0]{\begingroup\@sanitize@url \@url }%
\providecommand \@url [1]{\endgroup\@href {#1}{\urlprefix }}%
\providecommand \urlprefix  [0]{URL }%
\providecommand \Eprint [0]{\href }%
\providecommand \doibase [0]{https://doi.org/}%
\providecommand \selectlanguage [0]{\@gobble}%
\providecommand \bibinfo  [0]{\@secondoftwo}%
\providecommand \bibfield  [0]{\@secondoftwo}%
\providecommand \translation [1]{[#1]}%
\providecommand \BibitemOpen [0]{}%
\providecommand \bibitemStop [0]{}%
\providecommand \bibitemNoStop [0]{.\EOS\space}%
\providecommand \EOS [0]{\spacefactor3000\relax}%
\providecommand \BibitemShut  [1]{\csname bibitem#1\endcsname}%
\let\auto@bib@innerbib\@empty
\bibitem [{\citenamefont {{Neiman}}\ and\ \citenamefont {{Ward}}(1969)}]{NW}%
  \BibitemOpen
  \bibfield  {author} {\bibinfo {author} {\bibfnamefont {M.}~\bibnamefont {{Neiman}}}\ and\ \bibinfo {author} {\bibfnamefont {D.}~\bibnamefont {{Ward}}},\ }\href@noop {} {}\bibinfo {howpublished} {Beta decay half-lives for the new isotopes $^{166,164,162}${Lu}, $^{160,158}${Tm}, and $^{160,158}${Yb}, UCRL-18667 p. 59} (\bibinfo {year} {1969})\BibitemShut {NoStop}%
\bibitem [{ENS()}]{ENSDF}%
  \BibitemOpen
  \href@noop {} {}\bibinfo {howpublished} {Evaluated Nuclear Structure Data File (ENSDF) and XUNDL (October 2025 Version), continuously updated data files (NNDC, Brookhaven, NY)}\BibitemShut {NoStop}%
\bibitem [{\citenamefont {Schrewe}\ \emph {et~al.}(1982)\citenamefont {Schrewe}, \citenamefont {Hagberg}, \citenamefont {Schmeing}, \citenamefont {Hardy}, \citenamefont {Koslowsky}, \citenamefont {Sharma},\ and\ \citenamefont {Clifford}}]{Schrewe}%
  \BibitemOpen
  \bibfield  {author} {\bibinfo {author} {\bibfnamefont {U.~J.}\ \bibnamefont {Schrewe}}, \bibinfo {author} {\bibfnamefont {E.}~\bibnamefont {Hagberg}}, \bibinfo {author} {\bibfnamefont {H.}~\bibnamefont {Schmeing}}, \bibinfo {author} {\bibfnamefont {J.~C.}\ \bibnamefont {Hardy}}, \bibinfo {author} {\bibfnamefont {V.~T.}\ \bibnamefont {Koslowsky}}, \bibinfo {author} {\bibfnamefont {K.~S.}\ \bibnamefont {Sharma}},\ and\ \bibinfo {author} {\bibfnamefont {E.~T.~H.}\ \bibnamefont {Clifford}},\ }\bibfield  {title} {\bibinfo {title} {Decay studies of the new isotopes $^{162,163}${Hf}},\ }\href@noop {} {\bibfield  {journal} {\bibinfo  {journal} {Phys. Rev. C}\ }\textbf {\bibinfo {volume} {25}},\ \bibinfo {pages} {3091} (\bibinfo {year} {1982})}\BibitemShut {NoStop}%
\bibitem [{\citenamefont {Hild}\ \emph {et~al.}(1995)\citenamefont {Hild}, \citenamefont {Schmidt-Ott}, \citenamefont {Kunze}, \citenamefont {Meissner}, \citenamefont {Salewski}, \citenamefont {Toth},\ and\ \citenamefont {Michaelsen}}]{Hild162}%
  \BibitemOpen
  \bibfield  {author} {\bibinfo {author} {\bibfnamefont {T.}~\bibnamefont {Hild}}, \bibinfo {author} {\bibfnamefont {W.-D.}\ \bibnamefont {Schmidt-Ott}}, \bibinfo {author} {\bibfnamefont {V.}~\bibnamefont {Kunze}}, \bibinfo {author} {\bibfnamefont {F.}~\bibnamefont {Meissner}}, \bibinfo {author} {\bibfnamefont {H.}~\bibnamefont {Salewski}}, \bibinfo {author} {\bibfnamefont {K.~S.}\ \bibnamefont {Toth}},\ and\ \bibinfo {author} {\bibfnamefont {R.}~\bibnamefont {Michaelsen}},\ }\bibfield  {title} {\bibinfo {title} {Decays of $^{160-162}${Hf}},\ }\href@noop {} {\bibfield  {journal} {\bibinfo  {journal} {Phys. Rev. C}\ }\textbf {\bibinfo {volume} {52}},\ \bibinfo {pages} {223} (\bibinfo {year} {1995})}\BibitemShut {NoStop}%
\bibitem [{\citenamefont {Cardona}\ \emph {et~al.}(1997)\citenamefont {Cardona}, \citenamefont {Davidson}, \citenamefont {Hojman}, \citenamefont {Debray}, \citenamefont {Kreiner}, \citenamefont {Somacal}, \citenamefont {Davidson}, \citenamefont {Napoli}, \citenamefont {Bazzacco}, \citenamefont {Blasi}, \citenamefont {Burch}, \citenamefont {De~Acu\~na}, \citenamefont {Lenzi}, \citenamefont {Lo~Bianco}, \citenamefont {Rico},\ and\ \citenamefont {Rossi~Alvarez}}]{Cardona97}%
  \BibitemOpen
  \bibfield  {author} {\bibinfo {author} {\bibfnamefont {M.~A.}\ \bibnamefont {Cardona}}, \bibinfo {author} {\bibfnamefont {J.}~\bibnamefont {Davidson}}, \bibinfo {author} {\bibfnamefont {D.}~\bibnamefont {Hojman}}, \bibinfo {author} {\bibfnamefont {M.~E.}\ \bibnamefont {Debray}}, \bibinfo {author} {\bibfnamefont {A.~J.}\ \bibnamefont {Kreiner}}, \bibinfo {author} {\bibfnamefont {H.}~\bibnamefont {Somacal}}, \bibinfo {author} {\bibfnamefont {M.}~\bibnamefont {Davidson}}, \bibinfo {author} {\bibfnamefont {D.~R.}\ \bibnamefont {Napoli}}, \bibinfo {author} {\bibfnamefont {D.}~\bibnamefont {Bazzacco}}, \bibinfo {author} {\bibfnamefont {N.}~\bibnamefont {Blasi}}, \bibinfo {author} {\bibfnamefont {R.}~\bibnamefont {Burch}}, \bibinfo {author} {\bibfnamefont {D.}~\bibnamefont {De~Acu\~na}}, \bibinfo {author} {\bibfnamefont {S.~M.}\ \bibnamefont {Lenzi}}, \bibinfo {author} {\bibfnamefont {G.}~\bibnamefont {Lo~Bianco}}, \bibinfo {author} {\bibfnamefont {J.}~\bibnamefont {Rico}},\ and\ \bibinfo {author} {\bibfnamefont
  {C.}~\bibnamefont {Rossi~Alvarez}},\ }\bibfield  {title} {\bibinfo {title} {High-spin states in doubly odd $^{162,164}$ {Lu}},\ }\href@noop {} {\bibfield  {journal} {\bibinfo  {journal} {Phys. Rev. C}\ }\textbf {\bibinfo {volume} {56}},\ \bibinfo {pages} {707} (\bibinfo {year} {1997})}\BibitemShut {NoStop}%
\bibitem [{\citenamefont {Gupta}\ \emph {et~al.}(1997)\citenamefont {Gupta}, \citenamefont {Pancholi}, \citenamefont {Juneja}, \citenamefont {Mehta}, \citenamefont {Kumar}, \citenamefont {Bhowmik}, \citenamefont {Muralithar}, \citenamefont {Rodrigues},\ and\ \citenamefont {Singh}}]{Gupta97}%
  \BibitemOpen
  \bibfield  {author} {\bibinfo {author} {\bibfnamefont {S.~L.}\ \bibnamefont {Gupta}}, \bibinfo {author} {\bibfnamefont {S.~C.}\ \bibnamefont {Pancholi}}, \bibinfo {author} {\bibfnamefont {P.}~\bibnamefont {Juneja}}, \bibinfo {author} {\bibfnamefont {D.}~\bibnamefont {Mehta}}, \bibinfo {author} {\bibfnamefont {A.}~\bibnamefont {Kumar}}, \bibinfo {author} {\bibfnamefont {R.~K.}\ \bibnamefont {Bhowmik}}, \bibinfo {author} {\bibfnamefont {S.}~\bibnamefont {Muralithar}}, \bibinfo {author} {\bibfnamefont {G.}~\bibnamefont {Rodrigues}},\ and\ \bibinfo {author} {\bibfnamefont {R.~P.}\ \bibnamefont {Singh}},\ }\bibfield  {title} {\bibinfo {title} {High spin states in $^{162}${Lu}},\ }\href@noop {} {\bibfield  {journal} {\bibinfo  {journal} {Phys. Rev. C}\ }\textbf {\bibinfo {volume} {56}},\ \bibinfo {pages} {1281} (\bibinfo {year} {1997})}\BibitemShut {NoStop}%
\bibitem [{\citenamefont {Bringel}\ \emph {et~al.}(2003)\citenamefont {Bringel}, \citenamefont {Hübel}, \citenamefont {Amro}, \citenamefont {Axiotis}, \citenamefont {Bazzacco}, \citenamefont {Bhattacharya}, \citenamefont {Bhowmik}, \citenamefont {Domscheit}, \citenamefont {Hagemann}, \citenamefont {Jensen}, \citenamefont {Kröll}, \citenamefont {Lunardi}, \citenamefont {Napoli}, \citenamefont {Neußer}, \citenamefont {Pancholi}, \citenamefont {Petrache}, \citenamefont {Schönwaßer}, \citenamefont {Singh},\ and\ \citenamefont {Ur}}]{Bringel03}%
  \BibitemOpen
  \bibfield  {author} {\bibinfo {author} {\bibfnamefont {P.}~\bibnamefont {Bringel}}, \bibinfo {author} {\bibfnamefont {H.}~\bibnamefont {Hübel}}, \bibinfo {author} {\bibfnamefont {H.}~\bibnamefont {Amro}}, \bibinfo {author} {\bibfnamefont {M.}~\bibnamefont {Axiotis}}, \bibinfo {author} {\bibfnamefont {D.}~\bibnamefont {Bazzacco}}, \bibinfo {author} {\bibfnamefont {S.}~\bibnamefont {Bhattacharya}}, \bibinfo {author} {\bibfnamefont {R.}~\bibnamefont {Bhowmik}}, \bibinfo {author} {\bibfnamefont {J.}~\bibnamefont {Domscheit}}, \bibinfo {author} {\bibfnamefont {G.}~\bibnamefont {Hagemann}}, \bibinfo {author} {\bibfnamefont {D.}~\bibnamefont {Jensen}}, \bibinfo {author} {\bibfnamefont {T.}~\bibnamefont {Kröll}}, \bibinfo {author} {\bibfnamefont {S.}~\bibnamefont {Lunardi}}, \bibinfo {author} {\bibfnamefont {D.}~\bibnamefont {Napoli}}, \bibinfo {author} {\bibfnamefont {A.}~\bibnamefont {Neußer}}, \bibinfo {author} {\bibfnamefont {S.}~\bibnamefont {Pancholi}}, \bibinfo {author} {\bibfnamefont {C.}~\bibnamefont
  {Petrache}}, \bibinfo {author} {\bibfnamefont {G.}~\bibnamefont {Schönwaßer}}, \bibinfo {author} {\bibfnamefont {A.}~\bibnamefont {Singh}},\ and\ \bibinfo {author} {\bibfnamefont {C.}~\bibnamefont {Ur}},\ }\bibfield  {title} {\bibinfo {title} {First evidence for triaxial superdeformation in $^{161}${Lu} and $^{162}${Lu}},\ }\href@noop {} {\bibfield  {journal} {\bibinfo  {journal} {Eur. Phys. J.}\ }\textbf {\bibinfo {volume} {16}},\ \bibinfo {pages} {155} (\bibinfo {year} {2003})}\BibitemShut {NoStop}%
\bibitem [{\citenamefont {{\O}deg{\aa}rd}\ \emph {et~al.}(2001)\citenamefont {{\O}deg{\aa}rd}, \citenamefont {Hagemann}, \citenamefont {Jensen}, \citenamefont {Bergstr{\"o}m}, \citenamefont {Herskind}, \citenamefont {Sletten}, \citenamefont {T{\"o}rm{\"a}nen}, \citenamefont {Wilson}, \citenamefont {Tj{\o}m}, \citenamefont {Hamamoto}, \citenamefont {Spohr}, \citenamefont {H\"ubel}, \citenamefont {G\"orgen}, \citenamefont {Sch\"onwasser}, \citenamefont {Bracco}, \citenamefont {Leoni}, \citenamefont {Maj}, \citenamefont {Petrache}, \citenamefont {Bednarczyk},\ and\ \citenamefont {Curien}}]{163Lu}%
  \BibitemOpen
  \bibfield  {author} {\bibinfo {author} {\bibfnamefont {S.~W.}\ \bibnamefont {{\O}deg{\aa}rd}}, \bibinfo {author} {\bibfnamefont {G.~B.}\ \bibnamefont {Hagemann}}, \bibinfo {author} {\bibfnamefont {D.~R.}\ \bibnamefont {Jensen}}, \bibinfo {author} {\bibfnamefont {M.}~\bibnamefont {Bergstr{\"o}m}}, \bibinfo {author} {\bibfnamefont {B.}~\bibnamefont {Herskind}}, \bibinfo {author} {\bibfnamefont {G.}~\bibnamefont {Sletten}}, \bibinfo {author} {\bibfnamefont {S.}~\bibnamefont {T{\"o}rm{\"a}nen}}, \bibinfo {author} {\bibfnamefont {J.~N.}\ \bibnamefont {Wilson}}, \bibinfo {author} {\bibfnamefont {P.~O.}\ \bibnamefont {Tj{\o}m}}, \bibinfo {author} {\bibfnamefont {I.}~\bibnamefont {Hamamoto}}, \bibinfo {author} {\bibfnamefont {K.}~\bibnamefont {Spohr}}, \bibinfo {author} {\bibfnamefont {H.}~\bibnamefont {H\"ubel}}, \bibinfo {author} {\bibfnamefont {A.}~\bibnamefont {G\"orgen}}, \bibinfo {author} {\bibfnamefont {G.}~\bibnamefont {Sch\"onwasser}}, \bibinfo {author} {\bibfnamefont {A.}~\bibnamefont {Bracco}}, \bibinfo
  {author} {\bibfnamefont {S.}~\bibnamefont {Leoni}}, \bibinfo {author} {\bibfnamefont {A.}~\bibnamefont {Maj}}, \bibinfo {author} {\bibfnamefont {C.~M.}\ \bibnamefont {Petrache}}, \bibinfo {author} {\bibfnamefont {P.}~\bibnamefont {Bednarczyk}},\ and\ \bibinfo {author} {\bibfnamefont {D.}~\bibnamefont {Curien}},\ }\bibfield  {title} {\bibinfo {title} {Evidence for the wobbling mode in nuclei},\ }\href@noop {} {\bibfield  {journal} {\bibinfo  {journal} {Phys Rev. Lett.}\ }\textbf {\bibinfo {volume} {86}},\ \bibinfo {pages} {5866} (\bibinfo {year} {2001})}\BibitemShut {NoStop}%
\bibitem [{\citenamefont {{Bringel, P. and Hagemann, G. B. and Hübel, H. and Al-khatib, A. and Bednarczyk, P. and Bürger, A. and Curien, D. and Gangopadhyay, G. and Herskind, B. and Jensen, D. R. and Joss, D. T. and Kröll, Th. and Lo Bianco, G. and Lunardi, S. and Ma, W. C. and Nenoff, N. and Neußer-Neffgen, A. and Petrache, C. M. and Schönwasser, G. and Simpson, J. and Singh, A. K. and Singh, N. and Sletten, G.}}(2005)}]{161Lu}%
  \BibitemOpen
  \bibfield  {author} {\bibinfo {author} {\bibnamefont {{Bringel, P. and Hagemann, G. B. and Hübel, H. and Al-khatib, A. and Bednarczyk, P. and Bürger, A. and Curien, D. and Gangopadhyay, G. and Herskind, B. and Jensen, D. R. and Joss, D. T. and Kröll, Th. and Lo Bianco, G. and Lunardi, S. and Ma, W. C. and Nenoff, N. and Neußer-Neffgen, A. and Petrache, C. M. and Schönwasser, G. and Simpson, J. and Singh, A. K. and Singh, N. and Sletten, G.}}},\ }\bibfield  {title} {\bibinfo {title} {Evidence for wobbling excitation in $^{161}${Lu}},\ }\href@noop {} {\bibfield  {journal} {\bibinfo  {journal} {Eur. Phys. J. A}\ }\textbf {\bibinfo {volume} {24}},\ \bibinfo {pages} {167} (\bibinfo {year} {2005})}\BibitemShut {NoStop}%
\bibitem [{\citenamefont {Sugawara-Tanabe}\ and\ \citenamefont {Tanabe}(2012)}]{164LuTMT}%
  \BibitemOpen
  \bibfield  {author} {\bibinfo {author} {\bibfnamefont {K.}~\bibnamefont {Sugawara-Tanabe}}\ and\ \bibinfo {author} {\bibfnamefont {K.}~\bibnamefont {Tanabe}},\ }\bibfield  {title} {\bibinfo {title} {Tops-on-top model for triaxial, strongly deformed bands in $^{164}${Lu}},\ }\href@noop {} {\bibfield  {journal} {\bibinfo  {journal} {J. of Phys. Conf.}\ }\textbf {\bibinfo {volume} {381}},\ \bibinfo {pages} {012069} (\bibinfo {year} {2012})}\BibitemShut {NoStop}%
\bibitem [{\citenamefont {Törmänen}\ \emph {et~al.}(1999)\citenamefont {Törmänen}, \citenamefont {Ødegård}, \citenamefont {Hagemann}, \citenamefont {Harsmann}, \citenamefont {Bergström}, \citenamefont {Bark}, \citenamefont {Herskind}, \citenamefont {Sletten}, \citenamefont {Tjøm}, \citenamefont {Görgen}, \citenamefont {Hübel}, \citenamefont {Aengenvoort}, \citenamefont {{van Severen}}, \citenamefont {Fahlander}, \citenamefont {Napoli}, \citenamefont {Lenzi}, \citenamefont {Petrache}, \citenamefont {Ur}, \citenamefont {Jensen}, \citenamefont {Ryde}, \citenamefont {Bengtsson}, \citenamefont {Bracco}, \citenamefont {Frattini}, \citenamefont {Chapman}, \citenamefont {Cullen},\ and\ \citenamefont {King}}]{164Lutriax}%
  \BibitemOpen
  \bibfield  {author} {\bibinfo {author} {\bibfnamefont {S.}~\bibnamefont {Törmänen}}, \bibinfo {author} {\bibfnamefont {S.}~\bibnamefont {Ødegård}}, \bibinfo {author} {\bibfnamefont {G.}~\bibnamefont {Hagemann}}, \bibinfo {author} {\bibfnamefont {A.}~\bibnamefont {Harsmann}}, \bibinfo {author} {\bibfnamefont {M.}~\bibnamefont {Bergström}}, \bibinfo {author} {\bibfnamefont {R.}~\bibnamefont {Bark}}, \bibinfo {author} {\bibfnamefont {B.}~\bibnamefont {Herskind}}, \bibinfo {author} {\bibfnamefont {G.}~\bibnamefont {Sletten}}, \bibinfo {author} {\bibfnamefont {P.}~\bibnamefont {Tjøm}}, \bibinfo {author} {\bibfnamefont {A.}~\bibnamefont {Görgen}}, \bibinfo {author} {\bibfnamefont {H.}~\bibnamefont {Hübel}}, \bibinfo {author} {\bibfnamefont {B.}~\bibnamefont {Aengenvoort}}, \bibinfo {author} {\bibfnamefont {U.}~\bibnamefont {{van Severen}}}, \bibinfo {author} {\bibfnamefont {C.}~\bibnamefont {Fahlander}}, \bibinfo {author} {\bibfnamefont {D.}~\bibnamefont {Napoli}}, \bibinfo {author} {\bibfnamefont
  {S.}~\bibnamefont {Lenzi}}, \bibinfo {author} {\bibfnamefont {C.}~\bibnamefont {Petrache}}, \bibinfo {author} {\bibfnamefont {C.}~\bibnamefont {Ur}}, \bibinfo {author} {\bibfnamefont {H.}~\bibnamefont {Jensen}}, \bibinfo {author} {\bibfnamefont {H.}~\bibnamefont {Ryde}}, \bibinfo {author} {\bibfnamefont {R.}~\bibnamefont {Bengtsson}}, \bibinfo {author} {\bibfnamefont {A.}~\bibnamefont {Bracco}}, \bibinfo {author} {\bibfnamefont {S.}~\bibnamefont {Frattini}}, \bibinfo {author} {\bibfnamefont {R.}~\bibnamefont {Chapman}}, \bibinfo {author} {\bibfnamefont {D.}~\bibnamefont {Cullen}},\ and\ \bibinfo {author} {\bibfnamefont {S.}~\bibnamefont {King}},\ }\bibfield  {title} {\bibinfo {title} {Triaxial superdeformed bands in $^{164}${Lu} and enhanced e1 decay-out strength},\ }\href@noop {} {\bibfield  {journal} {\bibinfo  {journal} {Phys. Lett. B}\ }\textbf {\bibinfo {volume} {454}},\ \bibinfo {pages} {8} (\bibinfo {year} {1999})}\BibitemShut {NoStop}%
\bibitem [{\citenamefont {Schönwaßer}\ \emph {et~al.}(2003)\citenamefont {Schönwaßer}, \citenamefont {Hübel}, \citenamefont {Hagemann}, \citenamefont {Bednarczyk}, \citenamefont {Benzoni}, \citenamefont {Bracco}, \citenamefont {Bringel}, \citenamefont {Chapman}, \citenamefont {Curien}, \citenamefont {Domscheit}, \citenamefont {Herskind}, \citenamefont {Jensen}, \citenamefont {Leoni}, \citenamefont {{Lo Bianco}}, \citenamefont {Ma}, \citenamefont {Maj}, \citenamefont {Neußer}, \citenamefont {Ødegård}, \citenamefont {Petrache}, \citenamefont {Roßbach}, \citenamefont {Ryde}, \citenamefont {Spohr},\ and\ \citenamefont {Singh}}]{165Lu}%
  \BibitemOpen
  \bibfield  {author} {\bibinfo {author} {\bibfnamefont {G.}~\bibnamefont {Schönwaßer}}, \bibinfo {author} {\bibfnamefont {H.}~\bibnamefont {Hübel}}, \bibinfo {author} {\bibfnamefont {G.}~\bibnamefont {Hagemann}}, \bibinfo {author} {\bibfnamefont {P.}~\bibnamefont {Bednarczyk}}, \bibinfo {author} {\bibfnamefont {G.}~\bibnamefont {Benzoni}}, \bibinfo {author} {\bibfnamefont {A.}~\bibnamefont {Bracco}}, \bibinfo {author} {\bibfnamefont {P.}~\bibnamefont {Bringel}}, \bibinfo {author} {\bibfnamefont {R.}~\bibnamefont {Chapman}}, \bibinfo {author} {\bibfnamefont {D.}~\bibnamefont {Curien}}, \bibinfo {author} {\bibfnamefont {J.}~\bibnamefont {Domscheit}}, \bibinfo {author} {\bibfnamefont {B.}~\bibnamefont {Herskind}}, \bibinfo {author} {\bibfnamefont {D.}~\bibnamefont {Jensen}}, \bibinfo {author} {\bibfnamefont {S.}~\bibnamefont {Leoni}}, \bibinfo {author} {\bibfnamefont {G.}~\bibnamefont {{Lo Bianco}}}, \bibinfo {author} {\bibfnamefont {W.}~\bibnamefont {Ma}}, \bibinfo {author} {\bibfnamefont {A.}~\bibnamefont
  {Maj}}, \bibinfo {author} {\bibfnamefont {A.}~\bibnamefont {Neußer}}, \bibinfo {author} {\bibfnamefont {S.}~\bibnamefont {Ødegård}}, \bibinfo {author} {\bibfnamefont {C.}~\bibnamefont {Petrache}}, \bibinfo {author} {\bibfnamefont {D.}~\bibnamefont {Roßbach}}, \bibinfo {author} {\bibfnamefont {H.}~\bibnamefont {Ryde}}, \bibinfo {author} {\bibfnamefont {K.}~\bibnamefont {Spohr}},\ and\ \bibinfo {author} {\bibfnamefont {A.}~\bibnamefont {Singh}},\ }\bibfield  {title} {\bibinfo {title} {One- and two-phonon wobbling excitations in triaxial $^{165}${Lu}},\ }\href@noop {} {\bibfield  {journal} {\bibinfo  {journal} {Phys. Lett. B}\ }\textbf {\bibinfo {volume} {552}},\ \bibinfo {pages} {9} (\bibinfo {year} {2003})}\BibitemShut {NoStop}%
\bibitem [{\citenamefont {Amro}\ \emph {et~al.}(2003)\citenamefont {Amro}, \citenamefont {Ma}, \citenamefont {Hagemann}, \citenamefont {Diamond}, \citenamefont {Domscheit}, \citenamefont {Fallon}, \citenamefont {Görgen}, \citenamefont {Herskind}, \citenamefont {Hübel}, \citenamefont {Jensen}, \citenamefont {Li}, \citenamefont {Macchiavelli}, \citenamefont {Roux}, \citenamefont {Sletten}, \citenamefont {Thompson}, \citenamefont {Ward}, \citenamefont {Wiedenhöver}, \citenamefont {Wilson},\ and\ \citenamefont {Winger}}]{167Lu}%
  \BibitemOpen
  \bibfield  {author} {\bibinfo {author} {\bibfnamefont {H.}~\bibnamefont {Amro}}, \bibinfo {author} {\bibfnamefont {W.}~\bibnamefont {Ma}}, \bibinfo {author} {\bibfnamefont {G.}~\bibnamefont {Hagemann}}, \bibinfo {author} {\bibfnamefont {R.}~\bibnamefont {Diamond}}, \bibinfo {author} {\bibfnamefont {J.}~\bibnamefont {Domscheit}}, \bibinfo {author} {\bibfnamefont {P.}~\bibnamefont {Fallon}}, \bibinfo {author} {\bibfnamefont {A.}~\bibnamefont {Görgen}}, \bibinfo {author} {\bibfnamefont {B.}~\bibnamefont {Herskind}}, \bibinfo {author} {\bibfnamefont {H.}~\bibnamefont {Hübel}}, \bibinfo {author} {\bibfnamefont {D.}~\bibnamefont {Jensen}}, \bibinfo {author} {\bibfnamefont {Y.}~\bibnamefont {Li}}, \bibinfo {author} {\bibfnamefont {A.}~\bibnamefont {Macchiavelli}}, \bibinfo {author} {\bibfnamefont {D.}~\bibnamefont {Roux}}, \bibinfo {author} {\bibfnamefont {G.}~\bibnamefont {Sletten}}, \bibinfo {author} {\bibfnamefont {J.}~\bibnamefont {Thompson}}, \bibinfo {author} {\bibfnamefont {D.}~\bibnamefont {Ward}},
  \bibinfo {author} {\bibfnamefont {I.}~\bibnamefont {Wiedenhöver}}, \bibinfo {author} {\bibfnamefont {J.}~\bibnamefont {Wilson}},\ and\ \bibinfo {author} {\bibfnamefont {J.}~\bibnamefont {Winger}},\ }\bibfield  {title} {\bibinfo {title} {The wobbling mode in $^{167}${Lu}},\ }\href@noop {} {\bibfield  {journal} {\bibinfo  {journal} {Phys.Lett. B.}\ }\textbf {\bibinfo {volume} {553}},\ \bibinfo {pages} {197} (\bibinfo {year} {2003})}\BibitemShut {NoStop}%
\bibitem [{\citenamefont {Ya}\ \emph {et~al.}(2010)\citenamefont {Ya}, \citenamefont {Chen}, \citenamefont {Yu}, \citenamefont {Shen}, \citenamefont {Gao}, \citenamefont {Chen},\ and\ \citenamefont {Liu}}]{Lu_triax}%
  \BibitemOpen
  \bibfield  {author} {\bibinfo {author} {\bibfnamefont {T.}~\bibnamefont {Ya}}, \bibinfo {author} {\bibfnamefont {Y.}~\bibnamefont {Chen}}, \bibinfo {author} {\bibfnamefont {S.}~\bibnamefont {Yu}}, \bibinfo {author} {\bibfnamefont {C.}~\bibnamefont {Shen}}, \bibinfo {author} {\bibfnamefont {Z.}~\bibnamefont {Gao}}, \bibinfo {author} {\bibfnamefont {Y.}~\bibnamefont {Chen}},\ and\ \bibinfo {author} {\bibfnamefont {L.}~\bibnamefont {Liu}},\ }\bibfield  {title} {\bibinfo {title} {Triaxial superdeformed bands in odd–odd $^{160-168}${Lu} isotopes},\ }\href {https://doi.org/https://doi.org/10.1016/j.nuclphysa.2010.09.004} {\bibfield  {journal} {\bibinfo  {journal} {Nucl. Phys. A}\ }\textbf {\bibinfo {volume} {848}},\ \bibinfo {pages} {260} (\bibinfo {year} {2010})}\BibitemShut {NoStop}%
\bibitem [{\citenamefont {{Georg}}\ \emph {et~al.}(1998)\citenamefont {{Georg}}, \citenamefont {{W. Borchers}}, \citenamefont {{M. Keim}}, \citenamefont {{A. Klein}}, \citenamefont {{P. Lievens}}, \citenamefont {{R. Neugart}}, \citenamefont {{M. Neuroth}}, \citenamefont {{Pushpa M. Rao}}, \citenamefont {{Ch. Schulz}},\ and\ \citenamefont {{the ISOLDE Collaboration}}}]{Georg98}%
  \BibitemOpen
  \bibfield  {author} {\bibinfo {author} {\bibfnamefont {U.}~\bibnamefont {{Georg}}}, \bibinfo {author} {\bibnamefont {{W. Borchers}}}, \bibinfo {author} {\bibnamefont {{M. Keim}}}, \bibinfo {author} {\bibnamefont {{A. Klein}}}, \bibinfo {author} {\bibnamefont {{P. Lievens}}}, \bibinfo {author} {\bibnamefont {{R. Neugart}}}, \bibinfo {author} {\bibnamefont {{M. Neuroth}}}, \bibinfo {author} {\bibnamefont {{Pushpa M. Rao}}}, \bibinfo {author} {\bibnamefont {{Ch. Schulz}}},\ and\ \bibinfo {author} {\bibnamefont {{the ISOLDE Collaboration}}},\ }\bibfield  {title} {\bibinfo {title} {Laser spectroscopy investigation of the nuclear moments and radii of {Lutetium} isotopes},\ }\href@noop {} {\bibfield  {journal} {\bibinfo  {journal} {Eur. Phys. J. A.}\ }\textbf {\bibinfo {volume} {3}},\ \bibinfo {pages} {262} (\bibinfo {year} {1998})}\BibitemShut {NoStop}%
\bibitem [{\citenamefont {{Burman}}\ \emph {et~al.}(1978)\citenamefont {{Burman}}, \citenamefont {{Sen}},\ and\ \citenamefont {{Bakhru}}}]{Burman}%
  \BibitemOpen
  \bibfield  {author} {\bibinfo {author} {\bibfnamefont {C.}~\bibnamefont {{Burman}}}, \bibinfo {author} {\bibfnamefont {P.}~\bibnamefont {{Sen}}},\ and\ \bibinfo {author} {\bibfnamefont {H.}~\bibnamefont {{Bakhru}}},\ }\bibfield  {title} {\bibinfo {title} {Decay of $^{162,164,165}${Lu} isotopes},\ }\href@noop {} {\bibfield  {journal} {\bibinfo  {journal} {Can. J. Phys.}\ }\textbf {\bibinfo {volume} {56}},\ \bibinfo {pages} {786} (\bibinfo {year} {1978})}\BibitemShut {NoStop}%
\bibitem [{\citenamefont {Alkhazov}\ \emph {et~al.}(1979)\citenamefont {Alkhazov}, \citenamefont {Batist}, \citenamefont {Berlovich}, \citenamefont {Blinnikov}, \citenamefont {Yelkin}, \citenamefont {Mezilev}, \citenamefont {Novikov}, \citenamefont {Pantelejev}, \citenamefont {Poljakov}, \citenamefont {Schigolev}, \citenamefont {Tarasov}, \citenamefont {Afanasjev}, \citenamefont {Gromov}, \citenamefont {Jachim}, \citenamefont {Janicki}, \citenamefont {Kalinnikov}, \citenamefont {Kormicki}, \citenamefont {Potempa}, \citenamefont {Rurarz}, \citenamefont {Tarkanyi},\ and\ \citenamefont {Yushkievich}}]{Alkhazov}%
  \BibitemOpen
  \bibfield  {author} {\bibinfo {author} {\bibfnamefont {G.~D.}\ \bibnamefont {Alkhazov}}, \bibinfo {author} {\bibfnamefont {L.~K.}\ \bibnamefont {Batist}}, \bibinfo {author} {\bibfnamefont {E.~Y.}\ \bibnamefont {Berlovich}}, \bibinfo {author} {\bibfnamefont {Y.~S.}\ \bibnamefont {Blinnikov}}, \bibinfo {author} {\bibfnamefont {Y.~V.}\ \bibnamefont {Yelkin}}, \bibinfo {author} {\bibfnamefont {K.~A.}\ \bibnamefont {Mezilev}}, \bibinfo {author} {\bibfnamefont {Y.~N.}\ \bibnamefont {Novikov}}, \bibinfo {author} {\bibfnamefont {V.~N.}\ \bibnamefont {Pantelejev}}, \bibinfo {author} {\bibfnamefont {A.~G.}\ \bibnamefont {Poljakov}}, \bibinfo {author} {\bibfnamefont {N.~D.}\ \bibnamefont {Schigolev}}, \bibinfo {author} {\bibfnamefont {V.~N.}\ \bibnamefont {Tarasov}}, \bibinfo {author} {\bibfnamefont {V.~P.}\ \bibnamefont {Afanasjev}}, \bibinfo {author} {\bibfnamefont {K.~Y.}\ \bibnamefont {Gromov}}, \bibinfo {author} {\bibfnamefont {M.}~\bibnamefont {Jachim}}, \bibinfo {author} {\bibfnamefont {M.}~\bibnamefont
  {Janicki}}, \bibinfo {author} {\bibfnamefont {V.~G.}\ \bibnamefont {Kalinnikov}}, \bibinfo {author} {\bibfnamefont {J.}~\bibnamefont {Kormicki}}, \bibinfo {author} {\bibfnamefont {A.}~\bibnamefont {Potempa}}, \bibinfo {author} {\bibfnamefont {E.}~\bibnamefont {Rurarz}}, \bibinfo {author} {\bibfnamefont {F.}~\bibnamefont {Tarkanyi}},\ and\ \bibinfo {author} {\bibfnamefont {Y.~V.}\ \bibnamefont {Yushkievich}},\ }\bibfield  {title} {\bibinfo {title} {New neutron deficient $^{157,158,160,161,163}${Lu} isotopes},\ }\href@noop {} {\bibfield  {journal} {\bibinfo  {journal} {Zeitschrift f{\"u}r Physik A Atoms and Nuclei}\ }\textbf {\bibinfo {volume} {291}},\ \bibinfo {pages} {397} (\bibinfo {year} {1979})}\BibitemShut {NoStop}%
\bibitem [{\citenamefont {{Behrens}}(1980)}]{Behrens}%
  \BibitemOpen
  \bibfield  {author} {\bibinfo {author} {\bibfnamefont {H.}~\bibnamefont {{Behrens}}},\ }\emph {\bibinfo {title} {Kernspektroskopische Untersuchungen in $^{162}${Yb}}},\ \href@noop {} {Ph.D. thesis},\ \bibinfo  {school} {Tech. Univ. Munich} (\bibinfo {year} {1980})\BibitemShut {NoStop}%
\bibitem [{\citenamefont {{Nica}}(2024)}]{NDS}%
  \BibitemOpen
  \bibfield  {author} {\bibinfo {author} {\bibfnamefont {N.}~\bibnamefont {{Nica}}},\ }\bibfield  {title} {\bibinfo {title} {Nuclear data sheets for {A}=162},\ }\href@noop {} {\bibfield  {journal} {\bibinfo  {journal} {{Nucl. Dat. Sheets}}\ }\textbf {\bibinfo {volume} {195}},\ \bibinfo {pages} {1} (\bibinfo {year} {2024})}\BibitemShut {NoStop}%
\bibitem [{\citenamefont {Hunter}\ \emph {et~al.}(1977)\citenamefont {Hunter}, \citenamefont {Riedinger}, \citenamefont {Hillis}, \citenamefont {Bingham},\ and\ \citenamefont {Toth}}]{Hunt}%
  \BibitemOpen
  \bibfield  {author} {\bibinfo {author} {\bibfnamefont {R.~C.}\ \bibnamefont {Hunter}}, \bibinfo {author} {\bibfnamefont {L.~L.}\ \bibnamefont {Riedinger}}, \bibinfo {author} {\bibfnamefont {D.~L.}\ \bibnamefont {Hillis}}, \bibinfo {author} {\bibfnamefont {C.~R.}\ \bibnamefont {Bingham}},\ and\ \bibinfo {author} {\bibfnamefont {K.~S.}\ \bibnamefont {Toth}},\ }\bibfield  {title} {\bibinfo {title} {Levels in $^{164}\mathrm{Yb}$ from $^{164}\mathrm{Lu}$ decay},\ }\href@noop {} {\bibfield  {journal} {\bibinfo  {journal} {Phys. Rev. C}\ }\textbf {\bibinfo {volume} {16}},\ \bibinfo {pages} {384} (\bibinfo {year} {1977})}\BibitemShut {NoStop}%
\bibitem [{\citenamefont {{De Boer}}\ \emph {et~al.}(1974)\citenamefont {{De Boer}}, \citenamefont {Goudsmit}, \citenamefont {Koldewijn},\ and\ \citenamefont {Meyer}}]{Boer}%
  \BibitemOpen
  \bibfield  {author} {\bibinfo {author} {\bibfnamefont {F.}~\bibnamefont {{De Boer}}}, \bibinfo {author} {\bibfnamefont {P.}~\bibnamefont {Goudsmit}}, \bibinfo {author} {\bibfnamefont {P.}~\bibnamefont {Koldewijn}},\ and\ \bibinfo {author} {\bibfnamefont {B.}~\bibnamefont {Meyer}},\ }\bibfield  {title} {\bibinfo {title} {Decay properties of three $^{166}${Lu} isomers and the decay of $^{166}${Hf}},\ }\href@noop {} {\bibfield  {journal} {\bibinfo  {journal} {Nucl Phys. A}\ }\textbf {\bibinfo {volume} {225}},\ \bibinfo {pages} {317} (\bibinfo {year} {1974})}\BibitemShut {NoStop}%
\bibitem [{\citenamefont {Charvet}\ \emph {et~al.}(1972)\citenamefont {Charvet}, \citenamefont {Chéry}, \citenamefont {{Do Huu Phuoc}}, \citenamefont {Duffait}, \citenamefont {Emsallem},\ and\ \citenamefont {Marguier}}]{Charvet}%
  \BibitemOpen
  \bibfield  {author} {\bibinfo {author} {\bibfnamefont {A.}~\bibnamefont {Charvet}}, \bibinfo {author} {\bibfnamefont {R.}~\bibnamefont {Chéry}}, \bibinfo {author} {\bibnamefont {{Do Huu Phuoc}}}, \bibinfo {author} {\bibfnamefont {R.}~\bibnamefont {Duffait}}, \bibinfo {author} {\bibfnamefont {A.}~\bibnamefont {Emsallem}},\ and\ \bibinfo {author} {\bibfnamefont {G.}~\bibnamefont {Marguier}},\ }\bibfield  {title} {\bibinfo {title} {The decays of 5.5 min $^{168g}${Lu} and 6.7 min $^{168m}${Lu}},\ }\href@noop {} {\bibfield  {journal} {\bibinfo  {journal} {Nucl. Phys. A}\ }\textbf {\bibinfo {volume} {197}},\ \bibinfo {pages} {490} (\bibinfo {year} {1972})}\BibitemShut {NoStop}%
\bibitem [{\citenamefont {Barci}\ \emph {et~al.}(1997)\citenamefont {Barci}, \citenamefont {Ardisson}, \citenamefont {Trubert},\ and\ \citenamefont {Hussonnois}}]{Barci}%
  \BibitemOpen
  \bibfield  {author} {\bibinfo {author} {\bibfnamefont {V.}~\bibnamefont {Barci}}, \bibinfo {author} {\bibfnamefont {G.}~\bibnamefont {Ardisson}}, \bibinfo {author} {\bibfnamefont {D.}~\bibnamefont {Trubert}},\ and\ \bibinfo {author} {\bibfnamefont {M.}~\bibnamefont {Hussonnois}},\ }\bibfield  {title} {\bibinfo {title} {Excited states in the doubly odd $^{168}${Lu} nucleus fed by electron-capture decay of $^{168}${Hf} (t$_{1/2}$ =25.95 min)},\ }\href@noop {} {\bibfield  {journal} {\bibinfo  {journal} {Phys. Rev. C}\ }\textbf {\bibinfo {volume} {55}} (\bibinfo {year} {1997})}\BibitemShut {NoStop}%
\bibitem [{\citenamefont {Kardan}\ and\ \citenamefont {Sayyah}(2016)}]{ShapeLu}%
  \BibitemOpen
  \bibfield  {author} {\bibinfo {author} {\bibfnamefont {A.}~\bibnamefont {Kardan}}\ and\ \bibinfo {author} {\bibfnamefont {S.}~\bibnamefont {Sayyah}},\ }\bibfield  {title} {\bibinfo {title} {Shape evolution with angular momentum in {Lu} isotopes},\ }\href@noop {} {\bibfield  {journal} {\bibinfo  {journal} {Int. J. of Mod. Phys. E}\ }\textbf {\bibinfo {volume} {25}},\ \bibinfo {pages} {1650044} (\bibinfo {year} {2016})}\BibitemShut {NoStop}%
\bibitem [{\citenamefont {{R. Bengtsson and H. Ryde}}(2004)}]{Magicgaps}%
  \BibitemOpen
  \bibfield  {author} {\bibinfo {author} {\bibnamefont {{R. Bengtsson and H. Ryde}}},\ }\bibfield  {title} {\bibinfo {title} {Magic gaps and intruder levels in triaxially superdeformed nuclei},\ }\href@noop {} {\bibfield  {journal} {\bibinfo  {journal} {Eur. Phys. J. A}\ }\textbf {\bibinfo {volume} {22}},\ \bibinfo {pages} {355} (\bibinfo {year} {2004})}\BibitemShut {NoStop}%
\bibitem [{\citenamefont {Kumar}\ \emph {et~al.}(2011)\citenamefont {Kumar}, \citenamefont {Kumar}, \citenamefont {Kumar}, \citenamefont {Hasan}, \citenamefont {Koranga}, \citenamefont {Kumar}, \citenamefont {Negi},\ and\ \citenamefont {Angus}}]{stag}%
  \BibitemOpen
  \bibfield  {author} {\bibinfo {author} {\bibfnamefont {V.}~\bibnamefont {Kumar}}, \bibinfo {author} {\bibfnamefont {S.}~\bibnamefont {Kumar}}, \bibinfo {author} {\bibfnamefont {S.}~\bibnamefont {Kumar}}, \bibinfo {author} {\bibfnamefont {Z.}~\bibnamefont {Hasan}}, \bibinfo {author} {\bibfnamefont {B.~S.}\ \bibnamefont {Koranga}}, \bibinfo {author} {\bibfnamefont {D.}~\bibnamefont {Kumar}}, \bibinfo {author} {\bibfnamefont {D.}~\bibnamefont {Negi}},\ and\ \bibinfo {author} {\bibfnamefont {L.}~\bibnamefont {Angus}},\ }\bibfield  {title} {\bibinfo {title} {A semi-classical approach to signature splitting and signature inversion in odd-odd nuclei},\ }\href@noop {} {\bibfield  {journal} {\bibinfo  {journal} {Int. J. of Mod. Phys. E}\ }\textbf {\bibinfo {volume} {20}},\ \bibinfo {pages} {1455} (\bibinfo {year} {2011})}\BibitemShut {NoStop}%
\bibitem [{\citenamefont {Jain}\ \emph {et~al.}(1990)\citenamefont {Jain}, \citenamefont {Sheline}, \citenamefont {Sood},\ and\ \citenamefont {Jain}}]{Jain90}%
  \BibitemOpen
  \bibfield  {author} {\bibinfo {author} {\bibfnamefont {A.~K.}\ \bibnamefont {Jain}}, \bibinfo {author} {\bibfnamefont {R.~K.}\ \bibnamefont {Sheline}}, \bibinfo {author} {\bibfnamefont {P.~C.}\ \bibnamefont {Sood}},\ and\ \bibinfo {author} {\bibfnamefont {K.}~\bibnamefont {Jain}},\ }\bibfield  {title} {\bibinfo {title} {{Intrinsic States of Deformed Odd-{A} Nuclei in the Mass Regions (151 $\leq$ {A} $\leq$ 193) and ({A} $\geq$ 221)}},\ }\href@noop {} {\bibfield  {journal} {\bibinfo  {journal} {Rev. Mod. Phys.}\ }\textbf {\bibinfo {volume} {62}} (\bibinfo {year} {1990})}\BibitemShut {NoStop}%
\bibitem [{\citenamefont {{Headley}}\ \emph {et~al.}(1998)\citenamefont {{Headley}}, \citenamefont {{Sheline}}, \citenamefont {{Sood}}, \citenamefont {{Hoff}}, \citenamefont {{Hrĭ}vn{\`a}cov{\`a}}, \citenamefont {{Kvasil}}, \citenamefont {{Nosek}}, \citenamefont {{Jain}},\ and\ \citenamefont {{Burke}}}]{Headley98}%
  \BibitemOpen
  \bibfield  {author} {\bibinfo {author} {\bibfnamefont {D.~M.}\ \bibnamefont {{Headley}}}, \bibinfo {author} {\bibfnamefont {R.}~\bibnamefont {{Sheline}}}, \bibinfo {author} {\bibfnamefont {P.}~\bibnamefont {{Sood}}}, \bibinfo {author} {\bibfnamefont {R.}~\bibnamefont {{Hoff}}}, \bibinfo {author} {\bibfnamefont {I.}~\bibnamefont {{Hrĭ}vn{\`a}cov{\`a}}}, \bibinfo {author} {\bibfnamefont {J.}~\bibnamefont {{Kvasil}}}, \bibinfo {author} {\bibfnamefont {D.}~\bibnamefont {{Nosek}}}, \bibinfo {author} {\bibfnamefont {A.}~\bibnamefont {{Jain}}},\ and\ \bibinfo {author} {\bibfnamefont {D.}~\bibnamefont {{Burke}}},\ }\bibfield  {title} {\bibinfo {title} {{Intrinsic structures and associated rotational bands in medium-heavy deformed odd-odd nuclei}},\ }\href@noop {} {\bibfield  {journal} {\bibinfo  {journal} {At. Data Nucl. Data Tables}\ }\textbf {\bibinfo {volume} {69}},\ \bibinfo {pages} {239} (\bibinfo {year} {1998})}\BibitemShut {NoStop}%
\bibitem [{\citenamefont {Jain}\ \emph {et~al.}(1998)\citenamefont {Jain}, \citenamefont {Sheline}, \citenamefont {Headly}, \citenamefont {Sood}, \citenamefont {Burke}, \citenamefont {Hr\ifmmode~\breve{\imath}\else \u{\i}\fi{}vn\'acov\'a}, \citenamefont {Kvasil}, \citenamefont {Nosek},\ and\ \citenamefont {Hoff}}]{Jain98}%
  \BibitemOpen
  \bibfield  {author} {\bibinfo {author} {\bibfnamefont {A.~K.}\ \bibnamefont {Jain}}, \bibinfo {author} {\bibfnamefont {R.~K.}\ \bibnamefont {Sheline}}, \bibinfo {author} {\bibfnamefont {D.~M.}\ \bibnamefont {Headly}}, \bibinfo {author} {\bibfnamefont {P.~C.}\ \bibnamefont {Sood}}, \bibinfo {author} {\bibfnamefont {D.~G.}\ \bibnamefont {Burke}}, \bibinfo {author} {\bibfnamefont {I.}~\bibnamefont {Hr\ifmmode~\breve{\imath}\else \u{\i}\fi{}vn\'acov\'a}}, \bibinfo {author} {\bibfnamefont {J.}~\bibnamefont {Kvasil}}, \bibinfo {author} {\bibfnamefont {D.}~\bibnamefont {Nosek}},\ and\ \bibinfo {author} {\bibfnamefont {R.~W.}\ \bibnamefont {Hoff}},\ }\bibfield  {title} {\bibinfo {title} {Nuclear structure in odd-odd nuclei, 144$\leq$ {A} $\leq$194},\ }\href@noop {} {\bibfield  {journal} {\bibinfo  {journal} {Rev. Mod. Phys.}\ }\textbf {\bibinfo {volume} {70}},\ \bibinfo {pages} {843} (\bibinfo {year} {1998})}\BibitemShut {NoStop}%
\bibitem [{\citenamefont {Bohr}\ and\ \citenamefont {Mottelson}(1998)}]{BohrMottelson}%
  \BibitemOpen
  \bibfield  {author} {\bibinfo {author} {\bibfnamefont {A.}~\bibnamefont {Bohr}}\ and\ \bibinfo {author} {\bibfnamefont {B.}~\bibnamefont {Mottelson}},\ }\href@noop {} {\emph {\bibinfo {title} {Nuclear Structure Volume II: Deformations}}}\ (\bibinfo  {publisher} {World Scientific Publishing Co. Pte. Ltd.},\ \bibinfo {year} {1998})\BibitemShut {NoStop}%
\bibitem [{\citenamefont {{Gallagher}}\ and\ \citenamefont {{Moszkowski}}(1958)}]{GM}%
  \BibitemOpen
  \bibfield  {author} {\bibinfo {author} {\bibfnamefont {C.~J.}\ \bibnamefont {{Gallagher}}}\ and\ \bibinfo {author} {\bibfnamefont {S.~A.}\ \bibnamefont {{Moszkowski}}},\ }\bibfield  {title} {\bibinfo {title} {Coupling of angular momenta in odd-odd nuclei},\ }\href@noop {} {\bibfield  {journal} {\bibinfo  {journal} {Phys. Rev.}\ }\textbf {\bibinfo {volume} {111}},\ \bibinfo {pages} {1282} (\bibinfo {year} {1958})}\BibitemShut {NoStop}%
\bibitem [{\citenamefont {Sood}\ \emph {et~al.}(2021)\citenamefont {Sood}, \citenamefont {Jodidar},\ and\ \citenamefont {Gowrishankar}}]{tqrm252Md}%
  \BibitemOpen
  \bibfield  {author} {\bibinfo {author} {\bibfnamefont {P.~C.}\ \bibnamefont {Sood}}, \bibinfo {author} {\bibfnamefont {P.~M.}\ \bibnamefont {Jodidar}},\ and\ \bibinfo {author} {\bibfnamefont {R.}~\bibnamefont {Gowrishankar}},\ }\bibfield  {title} {\bibinfo {title} {Level structures of the transfermium odd-odd nucleus $^{252}${Md}},\ }\href@noop {} {\bibfield  {journal} {\bibinfo  {journal} {Phys. Rev. C}\ }\textbf {\bibinfo {volume} {103}} (\bibinfo {year} {2021})}\BibitemShut {NoStop}%
\bibitem [{\citenamefont {Sood}\ \emph {et~al.}(2012)\citenamefont {Sood}, \citenamefont {Gowrishankar},\ and\ \citenamefont {Sai}}]{tqrm154Pm}%
  \BibitemOpen
  \bibfield  {author} {\bibinfo {author} {\bibfnamefont {P.~C.}\ \bibnamefont {Sood}}, \bibinfo {author} {\bibfnamefont {R.}~\bibnamefont {Gowrishankar}},\ and\ \bibinfo {author} {\bibfnamefont {K.~V.}\ \bibnamefont {Sai}},\ }\bibfield  {title} {\bibinfo {title} {Level structures in the odd-odd nucleus $^{154}${Pm}},\ }\href@noop {} {\bibfield  {journal} {\bibinfo  {journal} {J. of Phys. G: Nucl. and Part. Phys.}\ }\textbf {\bibinfo {volume} {39}} (\bibinfo {year} {2012})}\BibitemShut {NoStop}%
\bibitem [{\citenamefont {{P. C. {Sood} and M. Sainath and R. Gowrishankar and B. Singh}}(2012)}]{tqrm156Pm}%
  \BibitemOpen
  \bibfield  {author} {\bibinfo {author} {\bibnamefont {{P. C. {Sood} and M. Sainath and R. Gowrishankar and B. Singh}}},\ }\bibfield  {title} {\bibinfo {title} {Level structures in $^{156}${Pm} from $^{156}${Nd} $\beta$-decay},\ }\href@noop {} {\bibfield  {journal} {\bibinfo  {journal} {Eur. Phys. J. A}\ }\textbf {\bibinfo {volume} {48}} (\bibinfo {year} {2012})}\BibitemShut {NoStop}%
\bibitem [{\citenamefont {{Gowrishankar}}\ and\ \citenamefont {{Sood}}(2016)}]{tqrm184Ta}%
  \BibitemOpen
  \bibfield  {author} {\bibinfo {author} {\bibfnamefont {R.}~\bibnamefont {{Gowrishankar}}}\ and\ \bibinfo {author} {\bibfnamefont {P.~C.}\ \bibnamefont {{Sood}}},\ }\bibfield  {title} {\bibinfo {title} {Level structures in odd-odd deformed nucleus $^{184}${Ta}},\ }\href@noop {} {\bibfield  {journal} {\bibinfo  {journal} {Eur. Phys. J. A}\ }\textbf {\bibinfo {volume} {52}} (\bibinfo {year} {2016})}\BibitemShut {NoStop}%
\bibitem [{\citenamefont {{Gowrishankar}}\ and\ \citenamefont {{Sood}}(2014)}]{tqrm186Ta}%
  \BibitemOpen
  \bibfield  {author} {\bibinfo {author} {\bibfnamefont {R.}~\bibnamefont {{Gowrishankar}}}\ and\ \bibinfo {author} {\bibfnamefont {P.~C.}\ \bibnamefont {{Sood}}},\ }\bibfield  {title} {\bibinfo {title} {Configuration assignments to isomers in the neutron-rich $^{186}${Ta} {(Z=73)} nucleus},\ }\href@noop {} {\bibfield  {journal} {\bibinfo  {journal} {Phys. Rev. C}\ }\textbf {\bibinfo {volume} {90}} (\bibinfo {year} {2014})}\BibitemShut {NoStop}%
\bibitem [{\citenamefont {Sood}\ \emph {et~al.}(2014)\citenamefont {Sood}, \citenamefont {Gowrishankar},\ and\ \citenamefont {Dora}}]{tqrm240Np}%
  \BibitemOpen
  \bibfield  {author} {\bibinfo {author} {\bibfnamefont {P.~C.}\ \bibnamefont {Sood}}, \bibinfo {author} {\bibfnamefont {R.}~\bibnamefont {Gowrishankar}},\ and\ \bibinfo {author} {\bibfnamefont {A.~K.}\ \bibnamefont {Dora}},\ }\bibfield  {title} {\bibinfo {title} {Level structures in $^{240}${Np}},\ }\href@noop {} {\bibfield  {journal} {\bibinfo  {journal} {Phys. Rev. C}\ }\textbf {\bibinfo {volume} {89}} (\bibinfo {year} {2014})}\BibitemShut {NoStop}%
\bibitem [{\citenamefont {{Sood}}\ and\ \citenamefont {R.{Gowrishankar}}(2022)}]{tqrm244Np}%
  \BibitemOpen
  \bibfield  {author} {\bibinfo {author} {\bibfnamefont {P.~C.}\ \bibnamefont {{Sood}}}\ and\ \bibinfo {author} {\bibnamefont {R.{Gowrishankar}}},\ }\bibfield  {title} {\bibinfo {title} {Low-lying level structures in the transuranic n-rich {Z}=93 nuclei $^{243}${Np} and $^{244}${Np}},\ }\href@noop {} {\bibfield  {journal} {\bibinfo  {journal} {Phys. Rev. C}\ }\textbf {\bibinfo {volume} {106}} (\bibinfo {year} {2022})}\BibitemShut {NoStop}%
\bibitem [{\citenamefont {Yadav}\ \emph {et~al.}(2014)\citenamefont {Yadav}, \citenamefont {Ma}, \citenamefont {Marsh}, \citenamefont {Ijaz}, \citenamefont {Janssens}, \citenamefont {Carpenter}, \citenamefont {Hoffman}, \citenamefont {Lauritsen}, \citenamefont {Zhu}, \citenamefont {Kondev}, \citenamefont {G\"urdal}, \citenamefont {Hagemann}, \citenamefont {Hartley}, \citenamefont {Riedinger},\ and\ \citenamefont {Mukhopadhyay}}]{Yadav}%
  \BibitemOpen
  \bibfield  {author} {\bibinfo {author} {\bibfnamefont {R.}~\bibnamefont {Yadav}}, \bibinfo {author} {\bibfnamefont {W.}~\bibnamefont {Ma}}, \bibinfo {author} {\bibfnamefont {J.}~\bibnamefont {Marsh}}, \bibinfo {author} {\bibfnamefont {Q.}~\bibnamefont {Ijaz}}, \bibinfo {author} {\bibfnamefont {R.}~\bibnamefont {Janssens}}, \bibinfo {author} {\bibfnamefont {M.}~\bibnamefont {Carpenter}}, \bibinfo {author} {\bibfnamefont {C.}~\bibnamefont {Hoffman}}, \bibinfo {author} {\bibfnamefont {T.}~\bibnamefont {Lauritsen}}, \bibinfo {author} {\bibfnamefont {S.}~\bibnamefont {Zhu}}, \bibinfo {author} {\bibfnamefont {F.}~\bibnamefont {Kondev}}, \bibinfo {author} {\bibfnamefont {G.}~\bibnamefont {G\"urdal}}, \bibinfo {author} {\bibfnamefont {G.}~\bibnamefont {Hagemann}}, \bibinfo {author} {\bibfnamefont {D.}~\bibnamefont {Hartley}}, \bibinfo {author} {\bibfnamefont {L.}~\bibnamefont {Riedinger}},\ and\ \bibinfo {author} {\bibfnamefont {S.}~\bibnamefont {Mukhopadhyay}},\ }\bibfield  {title} {\bibinfo {title} {Multiple
  excitation modes in $^{163}${Hf}},\ }\href@noop {} {\bibfield  {journal} {\bibinfo  {journal} {Phys. Rev. C}\ }\textbf {\bibinfo {volume} {90}},\ \bibinfo {pages} {054325} (\bibinfo {year} {2014})}\BibitemShut {NoStop}%
\bibitem [{\citenamefont {G.{Alaga}}(1957)}]{Alaga1957}%
  \BibitemOpen
  \bibfield  {author} {\bibinfo {author} {\bibnamefont {G.{Alaga}}},\ }\bibfield  {title} {\bibinfo {title} {Selection rules for beta and gamma particle transitions in strongly deformed nuclei},\ }\href@noop {} {\bibfield  {journal} {\bibinfo  {journal} {Nucl. Phys.}\ }\textbf {\bibinfo {volume} {4}},\ \bibinfo {pages} {625} (\bibinfo {year} {1957})}\BibitemShut {NoStop}%
\bibitem [{\citenamefont {Garg}\ \emph {et~al.}(2023)\citenamefont {Garg}, \citenamefont {Maheshwari}, \citenamefont {Singh}, \citenamefont {Sun}, \citenamefont {Goel},\ and\ \citenamefont {Jain}}]{Atlas}%
  \BibitemOpen
  \bibfield  {author} {\bibinfo {author} {\bibfnamefont {S.}~\bibnamefont {Garg}}, \bibinfo {author} {\bibfnamefont {B.}~\bibnamefont {Maheshwari}}, \bibinfo {author} {\bibfnamefont {B.}~\bibnamefont {Singh}}, \bibinfo {author} {\bibfnamefont {Y.}~\bibnamefont {Sun}}, \bibinfo {author} {\bibfnamefont {A.}~\bibnamefont {Goel}},\ and\ \bibinfo {author} {\bibfnamefont {A.~K.}\ \bibnamefont {Jain}},\ }\bibfield  {title} {\bibinfo {title} {Atlas of nuclear isomers-second edition},\ }\href@noop {} {\bibfield  {journal} {\bibinfo  {journal} {At. Data Nucl. Data Tables}\ }\textbf {\bibinfo {volume} {150}} (\bibinfo {year} {2023})}\BibitemShut {NoStop}%
\bibitem [{\citenamefont {{Sood}}\ and\ \citenamefont {{Sheline}}(1989)}]{SS89sf}%
  \BibitemOpen
  \bibfield  {author} {\bibinfo {author} {\bibfnamefont {P.~C.}\ \bibnamefont {{Sood}}}\ and\ \bibinfo {author} {\bibfnamefont {R.~K.}\ \bibnamefont {{Sheline}}},\ }\bibfield  {title} {\bibinfo {title} {Allowed spin-flip beta decays in medium-heavy (147 \textless {A} \textless 190) nuclei},\ }\href@noop {} {\bibfield  {journal} {\bibinfo  {journal} {At. Data Nucl. Data Tables}\ }\textbf {\bibinfo {volume} {43}},\ \bibinfo {pages} {259} (\bibinfo {year} {1989})}\BibitemShut {NoStop}%
\bibitem [{\citenamefont {{Sood}}\ and\ \citenamefont {{Sheline}}(1992)}]{SS92ext}%
  \BibitemOpen
  \bibfield  {author} {\bibinfo {author} {\bibfnamefont {P.~C.}\ \bibnamefont {{Sood}}}\ and\ \bibinfo {author} {\bibfnamefont {R.~K.}\ \bibnamefont {{Sheline}}},\ }\bibfield  {title} {\bibinfo {title} {Extended selection rules for allowed beta decays in actinides},\ }\href@noop {} {\bibfield  {journal} {\bibinfo  {journal} {Phys. Rev. C}\ }\textbf {\bibinfo {volume} {45}},\ \bibinfo {pages} {3006} (\bibinfo {year} {1992})}\BibitemShut {NoStop}%
\bibitem [{\citenamefont {{Gowrishankar}}\ \emph {et~al.}(2010)\citenamefont {{Gowrishankar}}, \citenamefont {{Vijay Sai}},\ and\ \citenamefont {P{.C. Sood}}}]{DAE1ff}%
  \BibitemOpen
  \bibfield  {author} {\bibinfo {author} {\bibfnamefont {R.}~\bibnamefont {{Gowrishankar}}}, \bibinfo {author} {\bibfnamefont {K.}~\bibnamefont {{Vijay Sai}}},\ and\ \bibinfo {author} {\bibnamefont {P{.C. Sood}}},\ }\bibfield  {title} {\bibinfo {title} {Forbidden, but fast, $\beta$-transitions in lighter rare-earth nuclei},\ }\href@noop {} {\bibfield  {journal} {\bibinfo  {journal} {Proc. of the DAE Symp. on Nucl. Phys.}\ }\textbf {\bibinfo {volume} {55}} (\bibinfo {year} {2010})}\BibitemShut {NoStop}%
\bibitem [{\citenamefont {Singh}\ \emph {et~al.}(1998)\citenamefont {Singh}, \citenamefont {Rodriguez}, \citenamefont {Wong},\ and\ \citenamefont {Tuli}}]{singh98}%
  \BibitemOpen
  \bibfield  {author} {\bibinfo {author} {\bibfnamefont {B.}~\bibnamefont {Singh}}, \bibinfo {author} {\bibfnamefont {J.}~\bibnamefont {Rodriguez}}, \bibinfo {author} {\bibfnamefont {S.}~\bibnamefont {Wong}},\ and\ \bibinfo {author} {\bibfnamefont {J.}~\bibnamefont {Tuli}},\ }\bibfield  {title} {\bibinfo {title} {{Review of log ft values in $\beta$ decay}},\ }\href@noop {} {\bibfield  {journal} {\bibinfo  {journal} {Nucl. Data Sheets}\ }\textbf {\bibinfo {volume} {84}},\ \bibinfo {pages} {487} (\bibinfo {year} {1998})}\BibitemShut {NoStop}%
\bibitem [{\citenamefont {McCutchan}\ \emph {et~al.}(2004)\citenamefont {McCutchan}, \citenamefont {Zamfir}, \citenamefont {Caprio}, \citenamefont {Casten}, \citenamefont {Amro}, \citenamefont {Beausang}, \citenamefont {Brenner}, \citenamefont {Hecht}, \citenamefont {Hutter}, \citenamefont {Langdown}, \citenamefont {Meyer}, \citenamefont {Regan}, \citenamefont {Ressler},\ and\ \citenamefont {Yamamoto}}]{McCutchen}%
  \BibitemOpen
  \bibfield  {author} {\bibinfo {author} {\bibfnamefont {E.~A.}\ \bibnamefont {McCutchan}}, \bibinfo {author} {\bibfnamefont {N.~V.}\ \bibnamefont {Zamfir}}, \bibinfo {author} {\bibfnamefont {M.~A.}\ \bibnamefont {Caprio}}, \bibinfo {author} {\bibfnamefont {R.~F.}\ \bibnamefont {Casten}}, \bibinfo {author} {\bibfnamefont {H.}~\bibnamefont {Amro}}, \bibinfo {author} {\bibfnamefont {C.~W.}\ \bibnamefont {Beausang}}, \bibinfo {author} {\bibfnamefont {D.~S.}\ \bibnamefont {Brenner}}, \bibinfo {author} {\bibfnamefont {A.~A.}\ \bibnamefont {Hecht}}, \bibinfo {author} {\bibfnamefont {C.}~\bibnamefont {Hutter}}, \bibinfo {author} {\bibfnamefont {S.~D.}\ \bibnamefont {Langdown}}, \bibinfo {author} {\bibfnamefont {D.~A.}\ \bibnamefont {Meyer}}, \bibinfo {author} {\bibfnamefont {P.~H.}\ \bibnamefont {Regan}}, \bibinfo {author} {\bibfnamefont {J.~J.}\ \bibnamefont {Ressler}},\ and\ \bibinfo {author} {\bibfnamefont {A.~D.}\ \bibnamefont {Yamamoto}},\ }\bibfield  {title} {\bibinfo {title} {Low spin states in $^{162}${Yb}
  and the {X(5)} critical point symmetry},\ }\href@noop {} {\bibfield  {journal} {\bibinfo  {journal} {Phys. Rev. C}\ }\textbf {\bibinfo {volume} {69}} (\bibinfo {year} {2004})}\BibitemShut {NoStop}%
\bibitem [{\citenamefont {Sood}\ \emph {et~al.}(1988)\citenamefont {Sood}, \citenamefont {Sheline},\ and\ \citenamefont {Hoff}}]{Sood170Lu}%
  \BibitemOpen
  \bibfield  {author} {\bibinfo {author} {\bibfnamefont {P.~C.}\ \bibnamefont {Sood}}, \bibinfo {author} {\bibfnamefont {R.~K.}\ \bibnamefont {Sheline}},\ and\ \bibinfo {author} {\bibfnamefont {R.~W.}\ \bibnamefont {Hoff}},\ }\bibfield  {title} {\bibinfo {title} {Characterization of energy levels in the nucleus $^{170}${Lu}},\ }\href@noop {} {\bibfield  {journal} {\bibinfo  {journal} {Pramana - J. Phys.}\ }\textbf {\bibinfo {volume} {30}},\ \bibinfo {pages} {93} (\bibinfo {year} {1988})}\BibitemShut {NoStop}%
\bibitem [{\citenamefont {Hild}\ \emph {et~al.}(1989)\citenamefont {Hild}, \citenamefont {Schmidt-Ott}, \citenamefont {Freystein}, \citenamefont {Meissner}, \citenamefont {Runte}, \citenamefont {Salewski},\ and\ \citenamefont {Michaelsen}}]{Hild164}%
  \BibitemOpen
  \bibfield  {author} {\bibinfo {author} {\bibfnamefont {T.}~\bibnamefont {Hild}}, \bibinfo {author} {\bibfnamefont {W.-D.}\ \bibnamefont {Schmidt-Ott}}, \bibinfo {author} {\bibfnamefont {V.}~\bibnamefont {Freystein}}, \bibinfo {author} {\bibfnamefont {F.}~\bibnamefont {Meissner}}, \bibinfo {author} {\bibfnamefont {E.}~\bibnamefont {Runte}}, \bibinfo {author} {\bibfnamefont {H.}~\bibnamefont {Salewski}},\ and\ \bibinfo {author} {\bibfnamefont {R.}~\bibnamefont {Michaelsen}},\ }\bibfield  {title} {\bibinfo {title} {{Spectroscopic investigation of $^{164,168}$Ta, $^{166}$W and $^{164,165}$Hf}},\ }\href@noop {} {\bibfield  {journal} {\bibinfo  {journal} {Nucl. Phys. A}\ }\textbf {\bibinfo {volume} {492}},\ \bibinfo {pages} {237} (\bibinfo {year} {1989})}\BibitemShut {NoStop}%
\bibitem [{\citenamefont {{Ekstrom}}\ \emph {et~al.}(1974)\citenamefont {{Ekstrom}}, \citenamefont {W.~{Hogervorst}},\ and\ \citenamefont {{Wannberg}}}]{Ekstrom}%
  \BibitemOpen
  \bibfield  {author} {\bibinfo {author} {\bibfnamefont {C.}~\bibnamefont {{Ekstrom}}}, \bibinfo {author} {\bibfnamefont {S.~I.}\ \bibnamefont {W.~{Hogervorst}}},\ and\ \bibinfo {author} {\bibfnamefont {G.}~\bibnamefont {{Wannberg}}},\ }\bibfield  {title} {\bibinfo {title} {Nuclear spins of $^{165}${Lu}, $^{168m}${Lu} and $^{172}${Lu}},\ }\href@noop {} {\bibfield  {journal} {\bibinfo  {journal} {Phys. Scr.}\ }\textbf {\bibinfo {volume} {10}} (\bibinfo {year} {1974})}\BibitemShut {NoStop}%
\bibitem [{\citenamefont {Ha}\ \emph {et~al.}(1998)\citenamefont {Ha}, \citenamefont {Kim}, \citenamefont {Lee}, \citenamefont {Huh}, \citenamefont {Moon}, \citenamefont {Gono}, \citenamefont {Mitarai}, \citenamefont {Morikawa}, \citenamefont {Shibata}, \citenamefont {Watanabe}, \citenamefont {Miyake}, \citenamefont {Komatsu}, \citenamefont {Odahara}, \citenamefont {Ideguchi},\ and\ \citenamefont {Zhou}}]{Ha1998}%
  \BibitemOpen
  \bibfield  {author} {\bibinfo {author} {\bibfnamefont {J.~H.}\ \bibnamefont {Ha}}, \bibinfo {author} {\bibfnamefont {J.~C.}\ \bibnamefont {Kim}}, \bibinfo {author} {\bibfnamefont {C.~S.}\ \bibnamefont {Lee}}, \bibinfo {author} {\bibfnamefont {J.~Y.}\ \bibnamefont {Huh}}, \bibinfo {author} {\bibfnamefont {C.-B.}\ \bibnamefont {Moon}}, \bibinfo {author} {\bibfnamefont {Y.}~\bibnamefont {Gono}}, \bibinfo {author} {\bibfnamefont {S.}~\bibnamefont {Mitarai}}, \bibinfo {author} {\bibfnamefont {T.}~\bibnamefont {Morikawa}}, \bibinfo {author} {\bibfnamefont {M.}~\bibnamefont {Shibata}}, \bibinfo {author} {\bibfnamefont {H.}~\bibnamefont {Watanabe}}, \bibinfo {author} {\bibfnamefont {M.}~\bibnamefont {Miyake}}, \bibinfo {author} {\bibfnamefont {E.}~\bibnamefont {Komatsu}}, \bibinfo {author} {\bibfnamefont {A.}~\bibnamefont {Odahara}}, \bibinfo {author} {\bibfnamefont {E.}~\bibnamefont {Ideguchi}},\ and\ \bibinfo {author} {\bibfnamefont {X.~H.}\ \bibnamefont {Zhou}},\ }\bibfield  {title} {\bibinfo {title} {High-spin
  states in odd-odd $^{168}${Lu}},\ }\href@noop {} {\bibfield  {journal} {\bibinfo  {journal} {Eur. Phys. J. A}\ }\textbf {\bibinfo {volume} {1}},\ \bibinfo {pages} {247} (\bibinfo {year} {1998})}\BibitemShut {NoStop}%
\bibitem [{\citenamefont {Ha}\ \emph {et~al.}(2002)\citenamefont {Ha}, \citenamefont {Kim}, \citenamefont {Lee}, \citenamefont {Lee}, \citenamefont {Huh}, \citenamefont {Moon}, \citenamefont {Chae}, \citenamefont {Furuno}, \citenamefont {Komatsubara}, \citenamefont {Schizuma}, \citenamefont {Matsuura}, \citenamefont {Kato}, \citenamefont {Sadaki}, \citenamefont {Ishiyama}, \citenamefont {Gono}, \citenamefont {Morikawa}, \citenamefont {Mitarai}, \citenamefont {Shibata}, \citenamefont {Watanabe}, \citenamefont {Miyake}, \citenamefont {Komatsu}, \citenamefont {Odahara}, \citenamefont {Ideguchi},\ and\ \citenamefont {Zhou}}]{Ha2002}%
  \BibitemOpen
  \bibfield  {author} {\bibinfo {author} {\bibfnamefont {J.~H.}\ \bibnamefont {Ha}}, \bibinfo {author} {\bibfnamefont {J.~C.}\ \bibnamefont {Kim}}, \bibinfo {author} {\bibfnamefont {C.~S.}\ \bibnamefont {Lee}}, \bibinfo {author} {\bibfnamefont {J.~H.}\ \bibnamefont {Lee}}, \bibinfo {author} {\bibfnamefont {J.~Y.}\ \bibnamefont {Huh}}, \bibinfo {author} {\bibfnamefont {C.-B.}\ \bibnamefont {Moon}}, \bibinfo {author} {\bibfnamefont {S.~J.}\ \bibnamefont {Chae}}, \bibinfo {author} {\bibfnamefont {K.}~\bibnamefont {Furuno}}, \bibinfo {author} {\bibfnamefont {T.}~\bibnamefont {Komatsubara}}, \bibinfo {author} {\bibfnamefont {T.}~\bibnamefont {Schizuma}}, \bibinfo {author} {\bibfnamefont {K.}~\bibnamefont {Matsuura}}, \bibinfo {author} {\bibfnamefont {K.}~\bibnamefont {Kato}}, \bibinfo {author} {\bibfnamefont {Y.}~\bibnamefont {Sadaki}}, \bibinfo {author} {\bibfnamefont {H.}~\bibnamefont {Ishiyama}}, \bibinfo {author} {\bibfnamefont {Y.}~\bibnamefont {Gono}}, \bibinfo {author} {\bibfnamefont {T.}~\bibnamefont
  {Morikawa}}, \bibinfo {author} {\bibfnamefont {S.}~\bibnamefont {Mitarai}}, \bibinfo {author} {\bibfnamefont {M.}~\bibnamefont {Shibata}}, \bibinfo {author} {\bibfnamefont {H.}~\bibnamefont {Watanabe}}, \bibinfo {author} {\bibfnamefont {M.}~\bibnamefont {Miyake}}, \bibinfo {author} {\bibfnamefont {E.}~\bibnamefont {Komatsu}}, \bibinfo {author} {\bibfnamefont {A.}~\bibnamefont {Odahara}}, \bibinfo {author} {\bibfnamefont {E.}~\bibnamefont {Ideguchi}},\ and\ \bibinfo {author} {\bibfnamefont {X.~H.}\ \bibnamefont {Zhou}},\ }\bibfield  {title} {\bibinfo {title} {High-spin states in odd--odd $^{168}${Lu}},\ }\href@noop {} {\bibfield  {journal} {\bibinfo  {journal} {J. Phys. Soc. of Japan}\ }\textbf {\bibinfo {volume} {71}},\ \bibinfo {pages} {1663} (\bibinfo {year} {2002})}\BibitemShut {NoStop}%
\bibitem [{\citenamefont {Roux}\ \emph {et~al.}(2021)\citenamefont {Roux}, \citenamefont {Ma}, \citenamefont {Hagemann}, \citenamefont {Elema}, \citenamefont {Fallon}, \citenamefont {G\"orgen}, \citenamefont {Herskind}, \citenamefont {H\"ubel}, \citenamefont {Marsh}, \citenamefont {Sletten}, \citenamefont {Ward},\ and\ \citenamefont {Wilson}}]{Roux2021}%
  \BibitemOpen
  \bibfield  {author} {\bibinfo {author} {\bibfnamefont {D.~G.}\ \bibnamefont {Roux}}, \bibinfo {author} {\bibfnamefont {W.~C.}\ \bibnamefont {Ma}}, \bibinfo {author} {\bibfnamefont {G.~B.}\ \bibnamefont {Hagemann}}, \bibinfo {author} {\bibfnamefont {D.~R.}\ \bibnamefont {Elema}}, \bibinfo {author} {\bibfnamefont {P.}~\bibnamefont {Fallon}}, \bibinfo {author} {\bibfnamefont {A.}~\bibnamefont {G\"orgen}}, \bibinfo {author} {\bibfnamefont {B.}~\bibnamefont {Herskind}}, \bibinfo {author} {\bibfnamefont {H.}~\bibnamefont {H\"ubel}}, \bibinfo {author} {\bibfnamefont {J.~C.}\ \bibnamefont {Marsh}}, \bibinfo {author} {\bibfnamefont {G.}~\bibnamefont {Sletten}}, \bibinfo {author} {\bibfnamefont {D.}~\bibnamefont {Ward}},\ and\ \bibinfo {author} {\bibfnamefont {J.~N.}\ \bibnamefont {Wilson}},\ }\bibfield  {title} {\bibinfo {title} {High-spin states and signature inversion in odd-odd $^{168}\mathrm{Lu}$},\ }\href {https://doi.org/10.1103/PhysRevC.103.034326} {\bibfield  {journal} {\bibinfo  {journal} {Phys. Rev. C}\
  }\textbf {\bibinfo {volume} {103}},\ \bibinfo {pages} {034326} (\bibinfo {year} {2021})}\BibitemShut {NoStop}%
\end{thebibliography}%

\end{document}